\documentclass[fleqn,usenatbib]{mnras}

\usepackage{afterpage}
\usepackage{newtxtext,newtxmath}
\usepackage{array}
\usepackage{multirow}

\usepackage{makecell}

\usepackage[T1]{fontenc}

\DeclareRobustCommand{\VAN}[3]{#2}
\let\VANthebibliography\thebibliography
\def\thebibliography{\DeclareRobustCommand{\VAN}[3]{##3}\VANthebibliography}

\usepackage{graphicx}	
\usepackage{amsmath}	
\usepackage[final]{changes}

\title[Prominence Parameters Effects on Lyman Lines]{Non-LTE Analysis of Pre-eruptive Prominence Plasma Parameters’ Effects on the Lyman $\beta$ and Lyman $\gamma$ Lines with Solar Orbiter SPICE Observations
}

\author[Y. Zhang et al.]{
Yong Zhang,$^{1}$
Nicolas Labrosse,$^{1}$
Susanna Parenti$^{2,3}$
and Therese A. Kucera$^{4}$
\\
$^{1}$SUPA School of Physics and Astronomy, University of Glasgow, Glasgow G12 8QQ UK\\
$^{2}$Institut d'Astrophysique Spatiale, Bat 121, Université Paris-Saclay/CNRS 91405 Orsay Cedex France\\
$^{3}$Universit\'e Paris-Saclay, Universit\'e Paris Cit\'e, CEA, CNRS, AIM, 91191, Gif-sur-Yvette, France\\
$^{4}$Heliophysics Science Division, NASA Goddard Space Flight Center, Greenbelt, MD, 20771 USA
}

\date{Accepted XXX. Received YYY; in original form ZZZ}

\pubyear{2015}

\begin{document}
\label{firstpage}
\pagerange{\pageref{firstpage}--\pageref{lastpage}}
\maketitle

\begin{abstract}
The first dedicated observation of an off-limb prominence by Solar Orbiter took place on April 15, 2023. Our aim is to determine the range of different physical parameters of this prominence and to examine how these parameters affect the formation of the Lyman $\beta$ and Lyman $\gamma$ lines of hydrogen. We have found a way to refine key physical parameters by observational data. We will test the method by this prominence observation. We generate 200 random non-LTE models using these observational constraints and compute the Lyman $\beta$ line and the Lyman $\gamma$ line profiles. We use the Spectral Imaging of the Coronal Environment (SPICE) full-disk mosaic from November 13, 2023 to constrain the incident radiation. We present the parameters and results of 200 random models using parallel coordinate plots to explore how different parameters affect the results. This allows us to infer the key physical parameters (e.g., central pressure, column mass and temperature gradient) that impact the formation of the Lyman $\beta$ line and the Lyman $\gamma$ line in this observation. 
\end{abstract}




\section{Introduction}

 The Solar Orbiter conducted its first dedicated off-limb prominence observation on April 15, 2023 \citep{muller2020solar}. Solar prominences, consisting of magnetically confined plasma suspended in the corona, represent key structures in solar atmospheric dynamics and are frequently associated with coronal mass ejections (CMEs), which have significant implications for space weather. Characterizing the physical conditions of prominences, such as temperature,  mass, pressure, and velocity, is essential for advancing solar physics and improving space weather forecasting.
 
In \cite{zhang2026analysis}, we presented the first comprehensive analysis of the Solar Orbiter/Spectral Imaging of the Coronal Environment (SPICE) off-limb prominence observation on April 15, 2023 \citep{2020A&A...642A..14S}. We investigated the spatial and temporal evolution of the H-Lyman~$\beta$ and H-Lyman~$\gamma$ line profiles in the pre-eruptive and eruptive phases of the prominence. By comparing integrated intensity and line width distributions between prominence, disk, and coronal regions, we revealed variations indicative of dynamic changes in density, temperature, and optical thickness. The study also introduced a geometric approach to derive radial velocities of eruptive filaments from a pair of H$\alpha$ images. Overall, the results highlighted the diagnostic potential of SPICE Lyman lines for probing prominence plasma conditions. However, \cite{zhang2026analysis} also highlighted that observational data alone cannot uniquely constrain key physical parameters, thereby motivating the need for detailed non-LTE radiative transfer modelling to fully interpret the observational data in this paper.

\cite{gouttebroze1993hydrogen} established 140 prominence models using 1D isothermal and isobaric slab geometries. Their systematic investigation provided comprehensive predictions of the hydrogen spectrum across a wide parameter space. This study revealed crucial correlations between plasma parameters and radiative outputs, particularly demonstrating a tight relationship between H$\alpha$ intensity and emission measure. \cite{warren1998high} provided observational constraints through high-resolution measurements of the Lyman series  (Ly$\beta$ to Ly$\lambda$) obtained with the SUMER instrument aboard SOHO. Their analysis revealed that Lyman lines show clear signatures of radiative transfer effects, with Ly$\beta$ through Ly$\epsilon$ exhibiting self-reversed profiles and higher series members displaying flat-topped shapes. These observations also documented significant center-to-limb variations and intensity-dependent profile asymmetries, providing essential validation for theoretical models and establishing reference profiles for incident radiation in prominence simulations.

More recently, \cite{gunar2020quiet} addressed the critical issue of temporal variability in incident radiation by deriving a reference quiet-Sun Lyman-$\alpha$ profile from SOHO/SUMER observations during the 2008 solar activity minimum. Their work provided not only a standardized profile for radiative transfer modeling but also a methodology to scale Lyman line intensities with the solar cycle using the LISIRD composite Lyman-$\alpha$ index. Through detailed analysis using a 2D prominence fine-structure model, they demonstrated that synthetic spectra are sensitive to changes in incident radiation, with central and integrated intensities of Lyman lines responding nearly proportionally to variations in solar illumination.

The primary objective of this study is to investigate how different plasma parameters (e.g., central pressure and central temperature) affect different properties (e.g., integrated intensity and optical thickness) of an eruptive solar prominence observed by SPICE on April 15, 2023, and to explore their influence on the formation of the Lyman $\beta$ and Lyman $\gamma$ lines of hydrogen. By analyzing the Lyman $\beta$ and Lyman $\gamma$ lines observed by SPICE, we aim to derive these key parameters and investigate their influence on the formation of Lyman lines. The data were taken during the pre-eruptive phase of the prominence. During this phase, plasma parameters such as temperature, density, and pressure may undergo significant changes as the prominence evolves toward eruption. For instance, an increase in magnetic stress or heating may lead to higher temperatures and enhanced line emissions, while changes in density distribution could affect the optical thickness and collisional contributions to the Lyman lines. Understanding these pre-eruptive conditions is crucial, as they set the stage for the subsequent eruption dynamics. This will provide valuable insights into the thermodynamic and dynamic properties of prominences eruption, contributing to a deeper understanding of their role in solar activity and space weather.

In Section~\ref{sec:obs}, we describe the instruments used in this work and provide an overview of the prominence observations for the event on April 15, 2023. In Section~\ref{sec:nlte}, we introduce the Non-LTE model used in this work, discuss the input of incident radiation and how we interpret model results via parallel coordinate plots and the elasticity coefficient. In Section~\ref{sec:effect}, we analyze the parameters' effect using parallel coordinate plots and introduce a method to refine parameters using the SPICE data. In Section~\ref{sec:discussion}, we discuss the results we obtain. In Section~\ref{sec:conclusion}, we summarize our conclusions and discuss the future work we plan to do.

\section{Observations}\label{sec:obs}
\subsection{Instruments}\label{sec:ins}

The Solar Orbiter, a solar observation mission developed by the European Space Agency (ESA) \citep{muller2020solar}, follows a highly elliptical orbit around the Sun. The majority of its remote sensing measurements are conducted near perihelion. At the time of this observation, Solar Orbiter was located at a heliocentric distance of 0.31~AU and positioned approximately in quadrature relative to Earth.


The SPICE instrument is an imaging spectrometer that operates in the extreme ultraviolet (EUV) range, covering 69.6–79.6~nm in the short-wavelength (SW) channel and 96.0–105.8~nm in the long-wavelength (LW) channel \citep{anderson2020solar}. Its primary mission is to support Solar Orbiter's scientific objectives by providing detailed diagnostics of the physical conditions and composition of plasma in the solar atmosphere.

\begin{figure}
    \centering
    \includegraphics[width=0.5\textwidth]{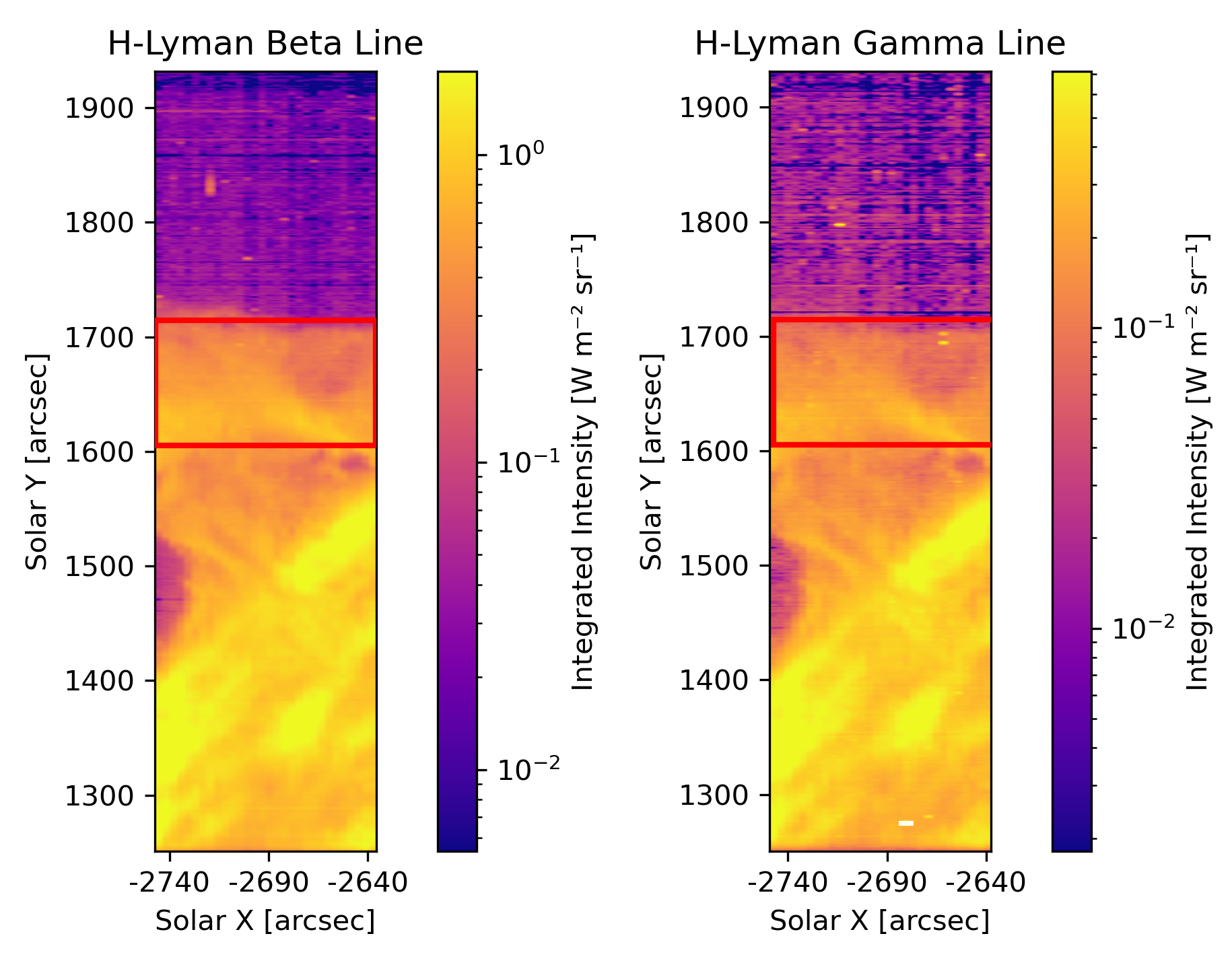}
    \caption{The SPICE observation of the integrated intensity of the Lyman $\beta$ and Lyman $\gamma$ lines taken from  07:03:40 - 08:09:23 UT, 15 April 2023. The red rectangle is the prominence region we study. }
    \label{fig:beta_gamma}
\end{figure}

SPICE plays a direct role in addressing Solar Orbiter's scientific objective related to coronal mass ejection (CME) formation by conducting coordinated prominence observations during 10-hour campaigns near perihelion. These campaigns utilize limb-pointing strategies in an Earth quadrature configuration, integrating SPICE’s spectral diagnostics with high-resolution imaging from the EUI instrument \citep{rochus2020solar}.




\added{In this paper, we focus on observations of Lyman $\beta$ and Lyman $\gamma$ lines. The rest wavelengths of Lyman $\beta$ and Lyman $\gamma$ lines are 102.57~nm and 97.25~nm respectively. The formation temperature of these lines is around a few $10^{4}~K$ \citep{dufresne2024chianti}. These lines are emitted by neutral hydrogen in the solar atmosphere and correspond to radiative transitions of neutral hydrogen from the $n=3$ and $n=4$ energy levels to the ground state ($n=1$), respectively. They are strong lines and well suited to study chromospheric plasmas, especially prominences }\citep{heinzel2001soho}. 

\subsection{Observations}\label{sec:obs2}

The prominence, a large quiescent structure located in the northern hemisphere, had been observable on the solar disk as seen from Earth for 6 days prior to this observation. Between 07:30 and 10:00 UT, it exhibited initial upward motion without undergoing eruption. At approximately 10:00 UT, the central segment of the prominence erupted, leaving residual material on both sides.

SPICE observations were conducted on 15 April 2023 from 01:03 to 14:58~UT. In this study, we utilize the pre-eruption rasters acquired between 07:03 and 08:09~UT to analyze the event (DOI: 10.48326/idoc.medoc.spice.5.0).


 Figure~\ref{fig:beta_gamma} presents SPICE observations of the prominence prior to eruption in the Lyman $\beta$ and Lyman $\gamma$ lines. Based on our previous work about observations \citep{zhang2026analysis}, the eruption is estimated to have commenced around 10:00~UT on 15 April 2023. Analysis of the SPICE data indicates that the prominence altitude ranged approximately from 20,000~km to 60,000~km during both the pre-eruption and eruption phases. \cite{zhang2026analysis} also showed that both lines exhibit strong intensity contrast between the solar disc, prominence, and corona, with the prominence displaying intermediate but spatially structured emission. The Lyman $\beta$ / Lyman $\gamma$ ratio remains relatively stable in the disc and prominence, while strong scatter appears in the corona due to weak signals.

\section{Modelling}\label{sec:nlte}

A non-LTE (non-local thermodynamic equilibrium) model is used to describe the physical properties of plasma in the solar atmosphere, where local thermodynamic equilibrium conditions are not satisfied. This approach enables the computation of various spectral profiles. For example, to determine the emergent intensities of the Lyman lines, the radiative transfer and statistical equilibrium equations for hydrogen are solved. These calculations assume the plasma to be in ionisation equilibrium and steady state, meaning that the population of each energy level of ions remains constant over time because of the balance between excitation and de-excitation processes. We will study the property of the pre-eruptive period of the prominence observed from 07:03 to 08:09~UT. We focus on this period because it is the only time period during which we have both Lyman $\beta$ and Lyman $\gamma$ lines observed simultaneously. During this period, the prominence has initial upward motion but the structure appears relatively stable. We consider the assumptions of steady-state and ionisation equilibrium to be a reasonable approximation. 

In this study, we use the 1D non-LTE radiative transfer code called PRODOP \citep{gouttebroze2000ready}. In this code, the eruptive prominence is modeled as a 1D slab oriented vertically above the solar surface, incorporating the prominence-corona transition region (PCTR). The model includes the radial velocity of the prominence as an input parameter. Additional input parameters include the incident spectral profile, the temperature and pressure at the slab center and at the boundary with the corona, the steepness of the temperature gradient in the PCTR, and the slab's total column mass, microturbulent velocity and altitude \citep{labrosse2012plasma}.  We use  temperature and pressure profiles from \cite{anzer1999energy} given by equations \ref{T} and \ref{p}: 

\begin{equation}
\begin{split}
{ T ( m ) = T _ { \text {cen} } + ( T _ { \text {tr} } - T _ { \text {cen} } ) ( 1 - 4 \frac { m } { M } ( 1 - \frac { m } { M } ) ) ^ { \gamma } } 
\end{split}
\label{T}
\end{equation}
\begin{equation}
\begin{split}
{ p ( m ) = 4 p _ { c } \frac { m } { M } ( 1 - \frac { m } { M } ) + p _ { \text {0} }  } 
\end{split}
\label{p}
\end{equation}

$T_{\text {cen}}$ is the central temperature and $T_{\text {tr}}$ is the boundary transition-region temperature. $\gamma$ is the steepness of the temperature gradient in the PCTR. $p_0$ is the pressure at the outer boundary and $p_{\text {cen}}=p_c+p_0$. $p_c$ is the difference between the pressure at the center and at the outer boundary. The column mass $m$ ranges from $m=0$ to $m=M$ (from one surface to the opposite surface). The computational methodology is detailed in \citet{labrosse2007effect, labrosse2008diagnostics}. We require estimates for the ranges of input parameters prior to inputting them into the model. Subsequently, the model can be used to investigate the formation of different spectral line profiles.

The source function of a spectral line is given by Equation~\ref{Eq: source}:
\begin{eqnarray}
S & = & (1 - \epsilon) \bar{J} + \epsilon B_{\nu_0}
\label{Eq: source}
\end{eqnarray}
In this equation, $\epsilon$ is the probability of a photon destruction. It represents the fraction of radiation that comes from thermal emission rather than scattering. $B_{\nu_0}$ in Eq. \ref{Eq: source} is the Plank function at frequency $\nu_0$. $\bar{J}$ is the mean intensity of the radiation field. $(1 - \epsilon) \bar{J}$ is the radiation term and $\epsilon B_{\nu_0}$ is the collision term.


\subsection{Incident radiation}
\label{Sec:incident}

An accurate determination of the radiation illuminating the structure is a requirement to study the formation of resonance lines in quiescent and eruptive prominences, as this incident radiation serves as boundary condition to solve the radiative transfer equation \citep[e.g.][]{labrosse2010,vial2015}. In the following we describe how we determine the incident radiation on the prominence analysed in this work by using SPICE full disc mosaic observations.

In order to have the intensity from the full Sun, we used data collected during the closest SPICE mosaic observation on Nov 13--14, 2023 to generate the incident Lyman $\beta$ line profile. This is the closest date of observation we can use, although it is still 7 months from the eruption. In Figure~\ref{fig:mosaic}, we show the data on the solar disk used to calculate the incident Lyman $\beta$ line profile. We follow the procedure of \cite{zhang2019prominences} to calculate the incident radiation. We have not considered the center-to-limb variation of the Lyman $\beta$ line because it is negligible. 

\begin{figure}
    \centering
\includegraphics[width=0.5\textwidth]{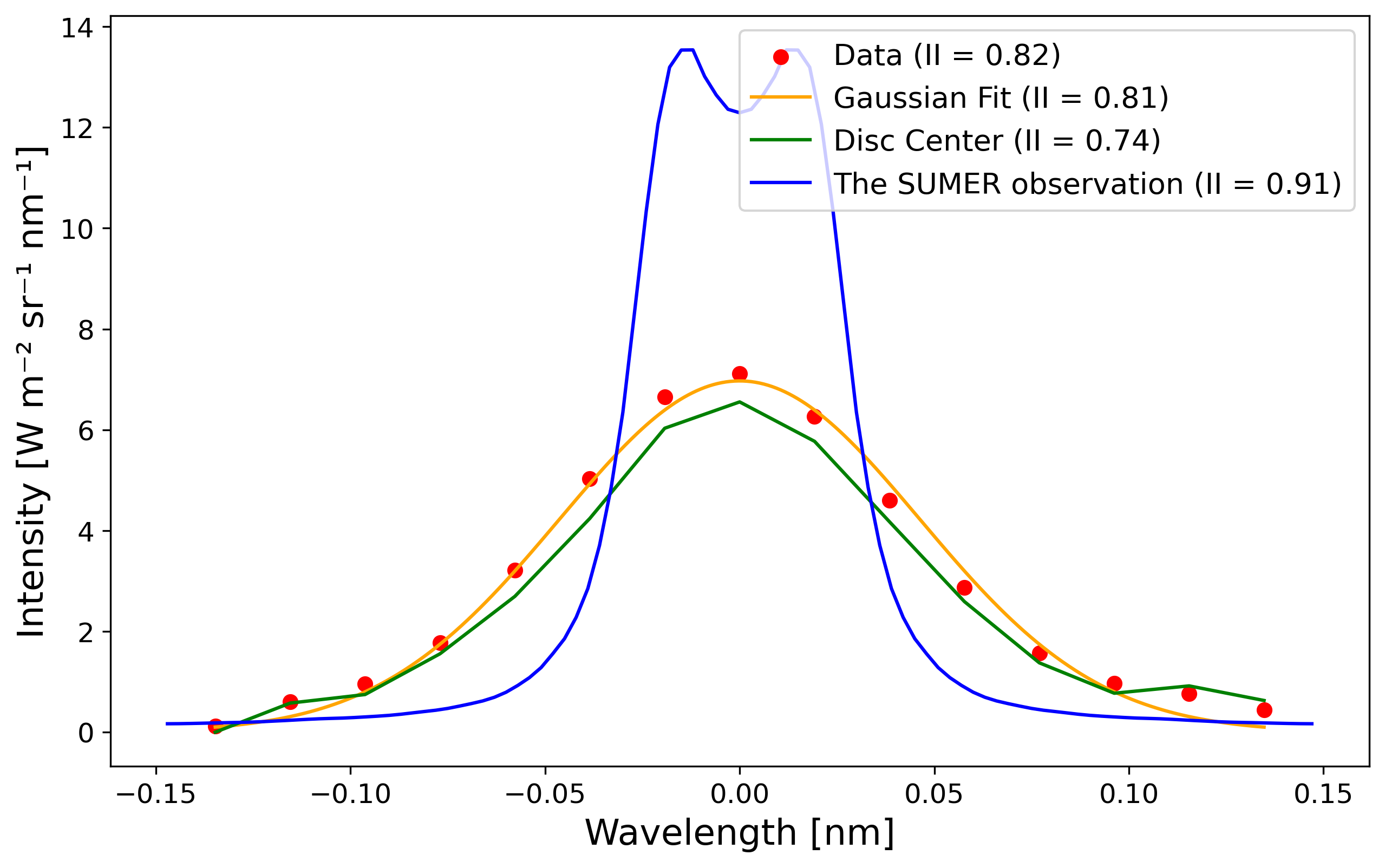}
\caption{
The red points are the incident Lyman $\beta$ line profile we obtained 
following the method of \protect\cite{zhang2019prominences} 
with SPICE full disc mosaic observation on Nov 13--14, 2023, 
and the orange line is the Gaussian fit of the red points. 
The green line is the average line profile within the 50 arcsec of the solar center in the mosaic image. 
The blue line is the incident Lyman $\beta$ line profile in the PRODOP NLTE code. 
It is obtained by the SUMER instrument on SOHO and presented in \protect\cite{warren1998high}. 
II is the abbreviation of Integrated Intensity, with units of W m$^{-2}$ sr$^{-1}$.
}
    \label{fig:compare}
\end{figure}

In Figure~\ref{fig:compare}, the red points are the incident Lyman $\beta$ line profile we obtain by integrating over the entire disk and the orange line is the Gaussian fit of the red points. The green line is the average line profile within the 50 arcsec of the solar center in the mosaic image. The blue line is the original Lyman $\beta$ line profile in the PRODOP NLTE code. It is a disk-averaged profile from \cite{warren1998high} and also the default input of the PRODOP NLTE code. 

However, we cannot use the incident Lyman $\beta$ line profile we obtained (orange line in Fig.~\ref{fig:compare}) directly, as the incident spectral profile is convolved with the broad SPICE instrumental profile. It is clear that the typical line shape of the Lyman $\beta$ line with its two peaks and central reversal is lost due to the convolution with the instrumental profile. Using this profile would not enable us to compute realistic emergent profiles. The SPICE Point Spread Function introduces several artifacts which are discussed in \cite{plowman2023spice}. For our purpose, we preferred to avoid the deconvolution of the profile and instead adopt the profile from \cite{warren1998high}. Therefore, we only use the integrated intensity of the observed incident Lyman $\beta$ line profile to re-scale the original Lyman $\beta$ line profile used in the NLTE code. We also normalize the original Lyman $\gamma$ line profile in the NLTE code by the same ratio we apply to normalize the Lyman $\beta$ line profile in the NLTE code. As we cannot determine more precisely the line profiles to determine the incident radiation on the day of these observations, it is reasonable to assume that the overall shapes of both lines does not vary significantly over time.

\begin{figure}
    \centering
\includegraphics[width=0.5\textwidth]{mosaic.png}
    \caption{The solar map of the integrated intensity of the \added{H-}Lyman $\beta$ line by SPICE full disc mosaic observation on Nov 13$\sim$14, 2023.}
    \label{fig:mosaic}
\end{figure}

\subsection{Toolkit to interpret model results}

\subsubsection{Parallel coordinate plot}
We use the method of the parallel coordinate plot to analyze the parameters' effect in the 2023 April 15th event \citep{edsall2003parallel,ge2009exploring}. Parallel coordinate plots provide a method to visualize relationships in multivariate data. With this method, we can see a parameter's effect more obviously without setting other parameters to specific values, and we can simultaneously display and compare values and relationships of multiple variables, even for non-linear relationships. We demonstrate the potential of this method by studying the well-known correlation between H$\alpha$ intensity and emission measure in prominences \citep{gouttebroze1993hydrogen, heinzel1994theoretical}. We create 200 random isobaric and isothermal models with the same parameters' ranges as \cite{heinzel1994theoretical}.

\begin{figure*}
    \centering
    \includegraphics[width=0.95\textwidth]{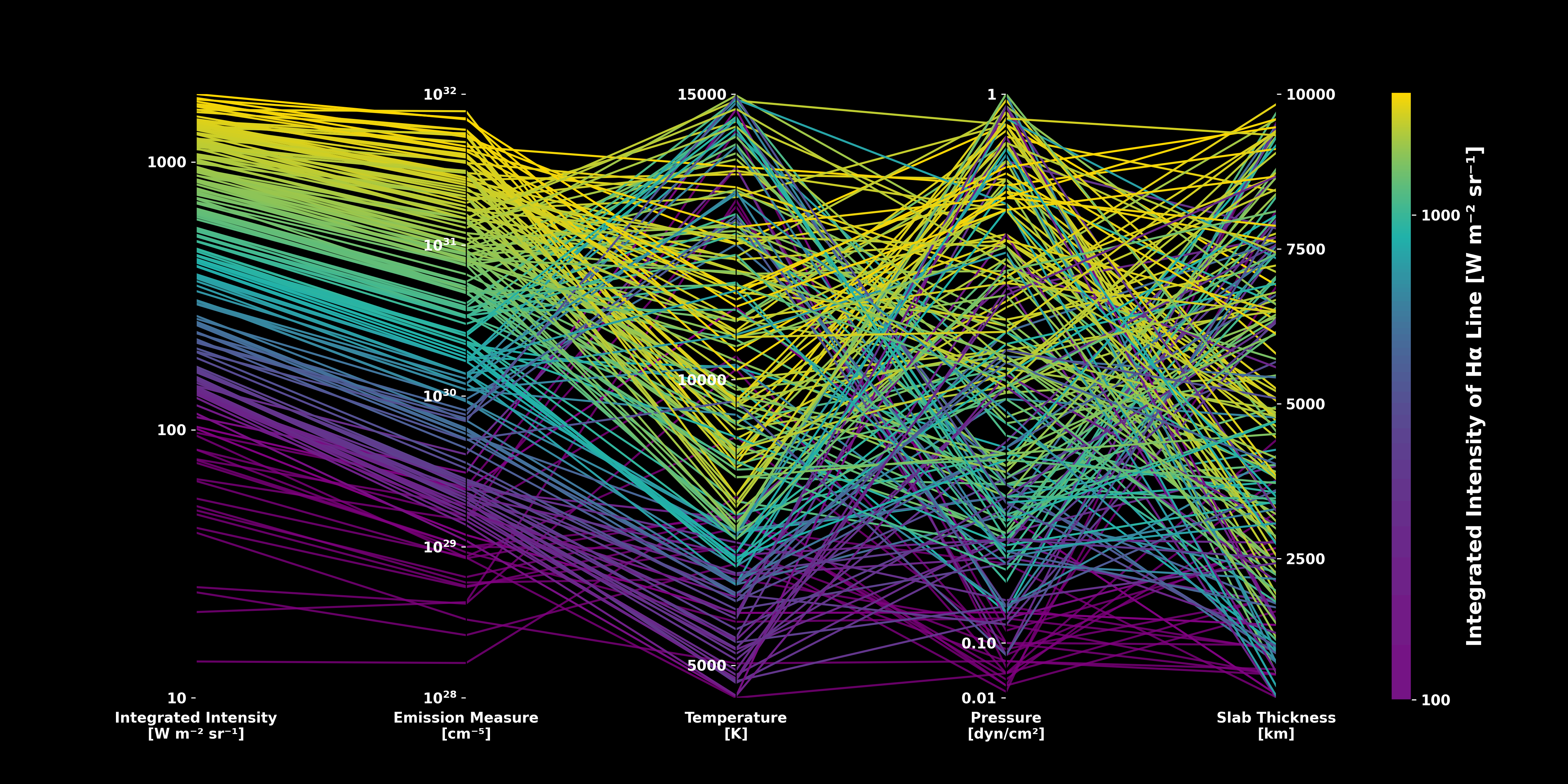}
    \caption{The parallel coordinate plot of 4 parameters with the integrated intensity of the H$\alpha$ line. In a parallel coordinate plot, if we see a diversity of colours connected to different value ranges when we focus on one parameter's axis, this means the parameter does not have much effect; on the other hand, if we see a certain colour concentrated on a certain value range, this means the parameter has some effect. For example, we can see that in the axis of emission measure, temperature, and pressure, the colour seems to be more uniform in different parts. This suggests that these parameters have relatively high correlation coefficients (usually larger than 0.3) with integrated intensity. }
    \label{fig:emission_measure}
\end{figure*}

\begin{figure*}
    \centering
    \includegraphics[width=\textwidth]{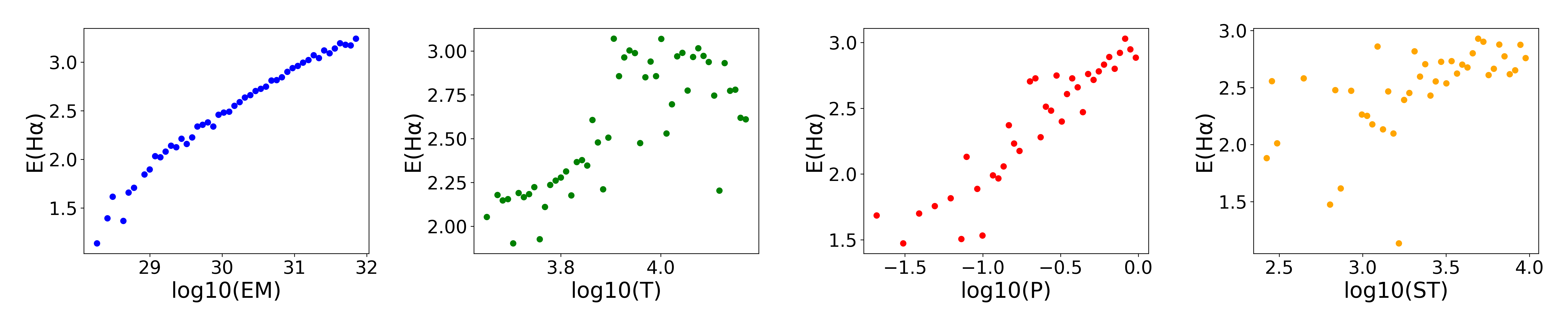}
    \caption{Integrated intensity vs. emission measure, temperature, pressure and slab thickness for the H$\alpha$ line. We divided the range of each x-axis parameter (log\,EM, log\,T, log\,p, log\,ST) into 50 bins. For each bin, all models falling inside it (the number varies across bins) were averaged in log\,E(H$\alpha$), and only bins containing data are shown.}
    \label{fig:Heinzel1994para}
\end{figure*}

\begin{figure*}
    \centering
    \includegraphics[width=0.95\textwidth]{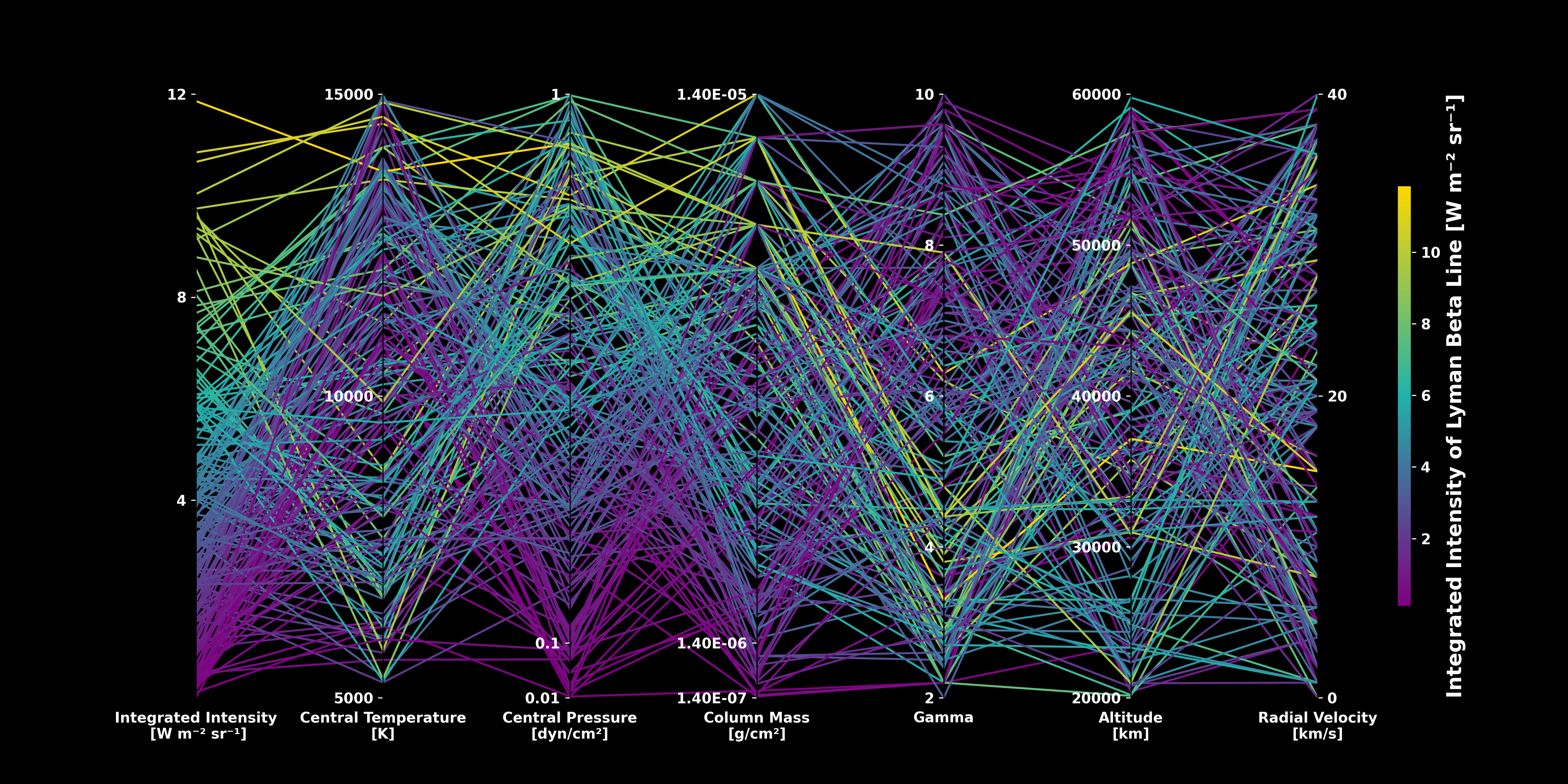}
    \caption{The parallel coordinate plot of 6 parameters with the integrated intensity of the Lyman $\beta$ line. }
    \label{fig:II1026}
\end{figure*}


\begin{figure*}
    \centering
    \includegraphics[width=0.95\textwidth]{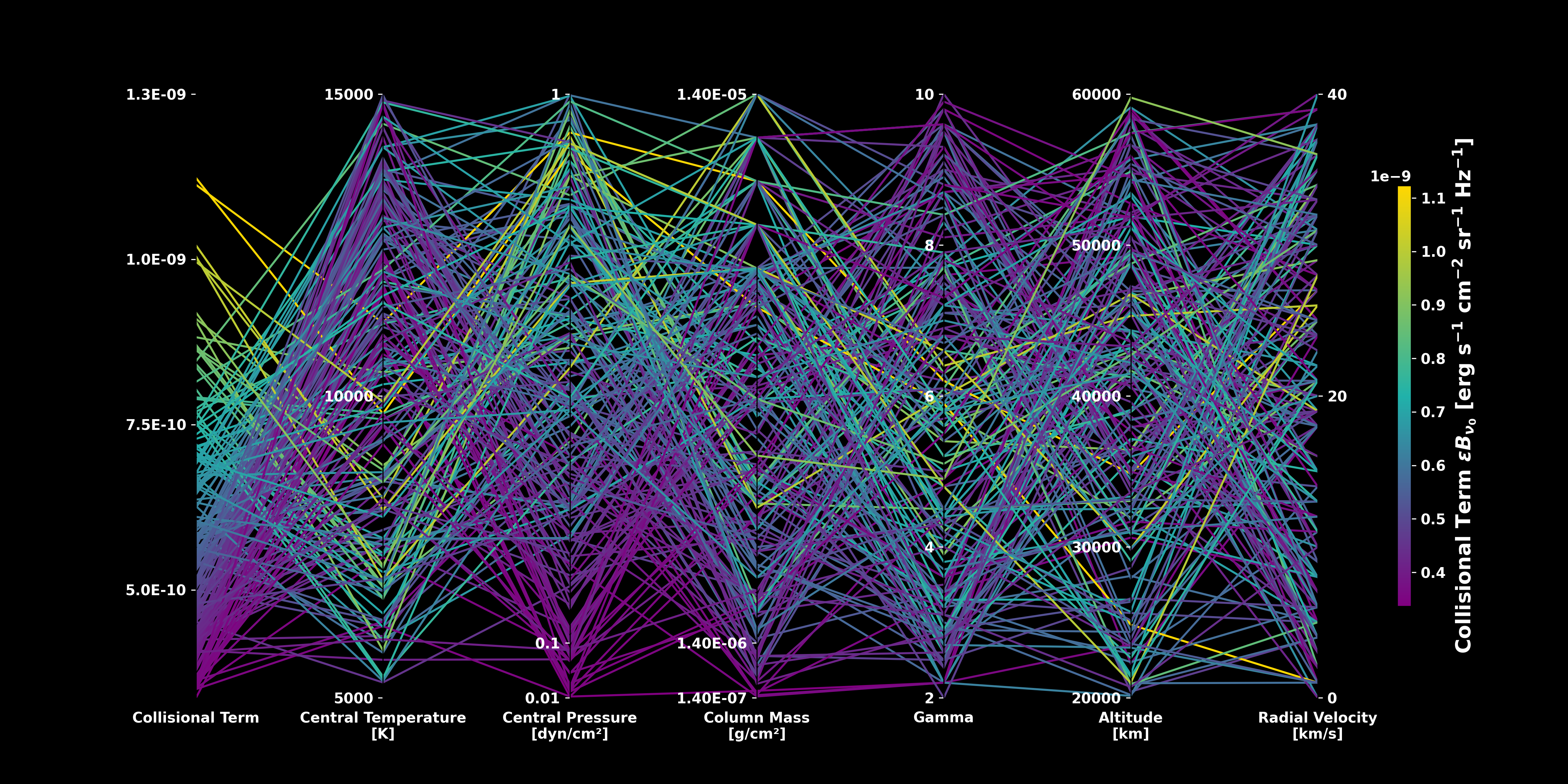}
    \caption{The parallel coordinate plot of 6 parameters with the collisional term $\epsilon B_{\nu_0}$ of the Lyman $\beta$ line.}
    \label{fig:CT1026}
\end{figure*}

\begin{figure*}
    \centering
    \includegraphics[width=0.95\textwidth]{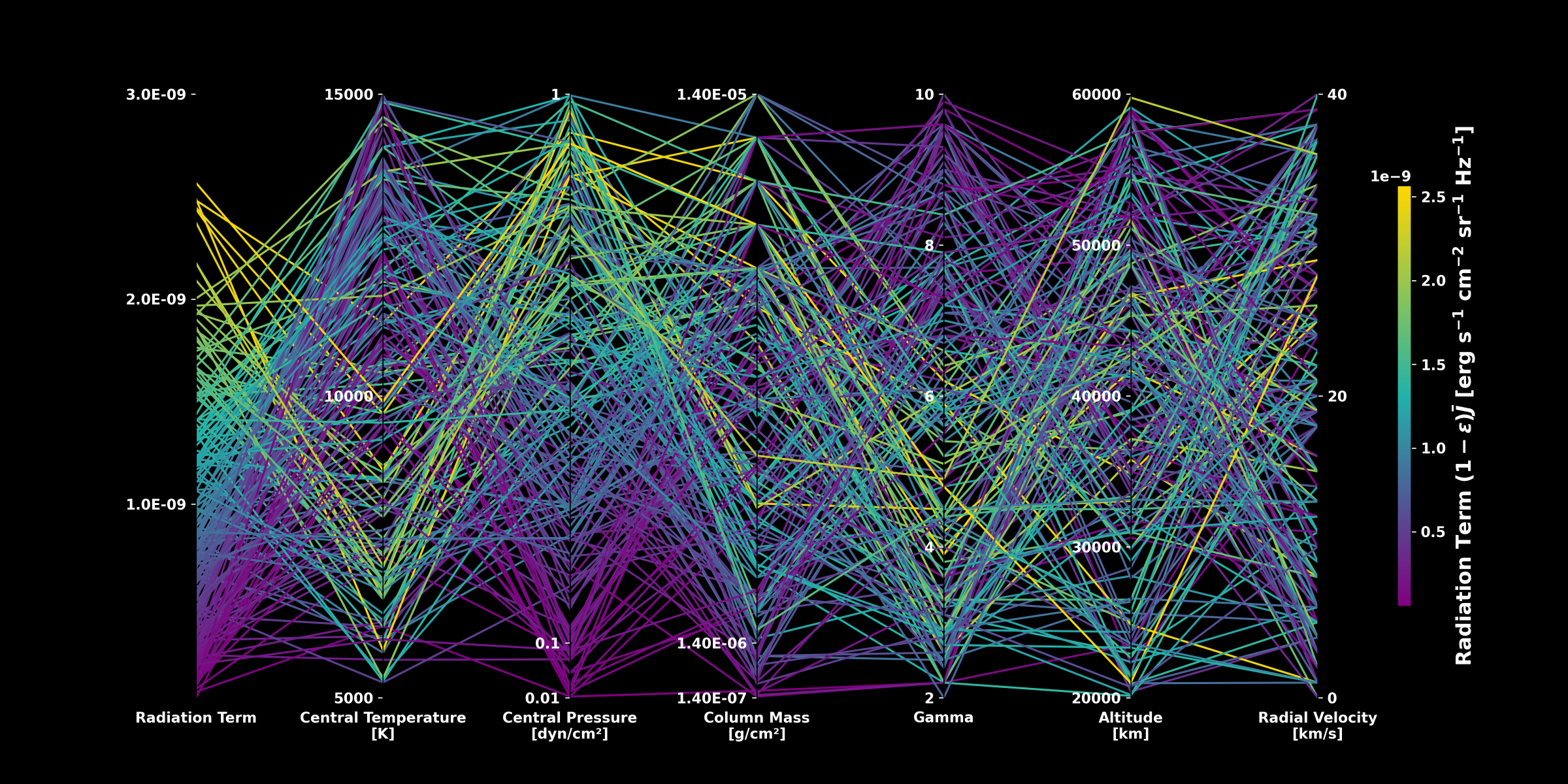}
    \caption{The parallel coordinate plot of 6 parameters with the radiation term $(1 - \epsilon) \bar{J}$ of the Lyman $\beta$ line.}
    \label{fig:RT1026}
\end{figure*}

\begin{figure*}
    \centering
    \includegraphics[width=0.95\textwidth]{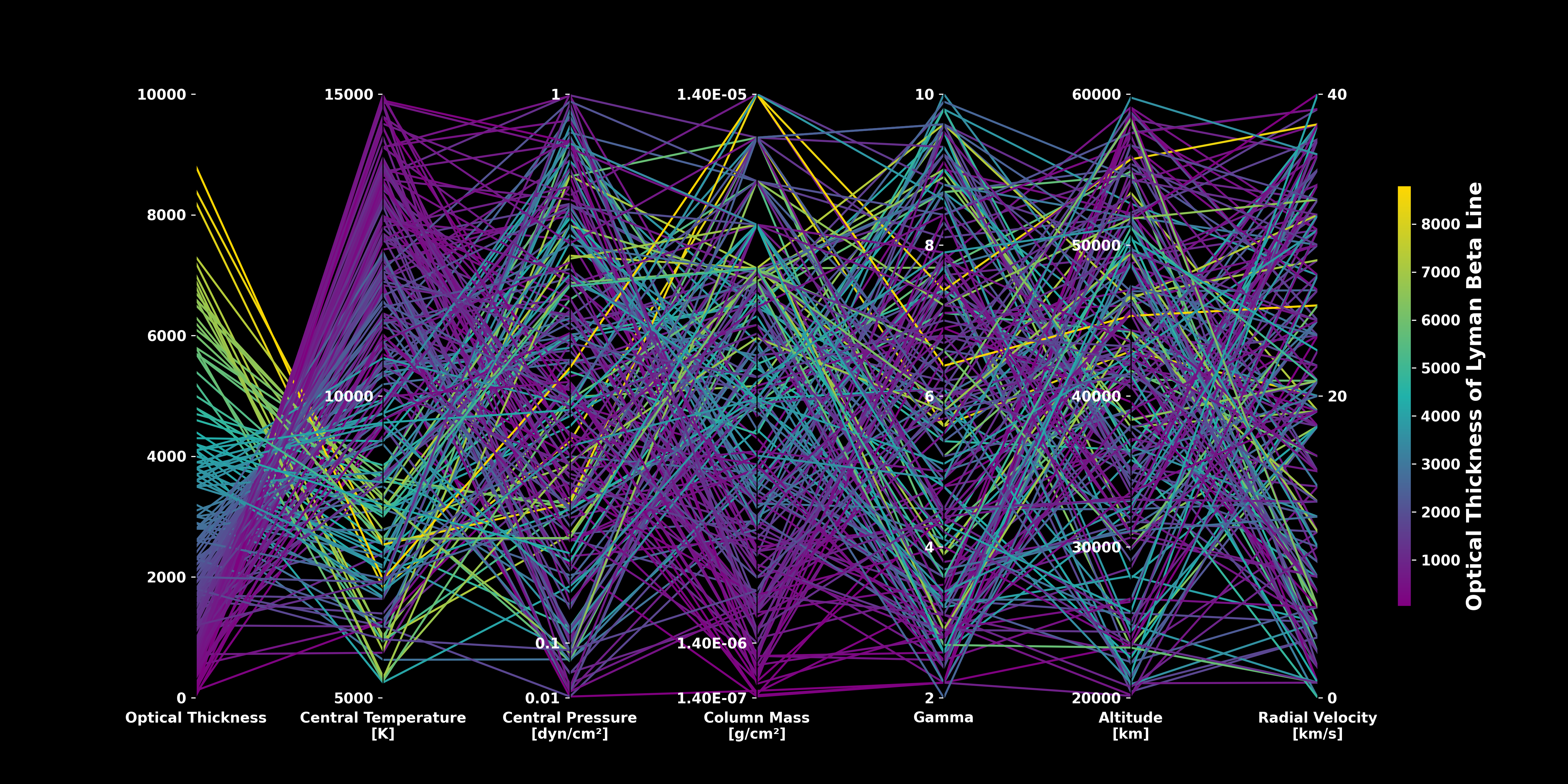}
    \caption{The parallel coordinate plot of 6 parameters with the optical thickness of the Lyman $\beta$ line.}
    \label{fig:OT1026}
\end{figure*}

\begin{figure*}
    \centering
    \includegraphics[width=0.95\textwidth]{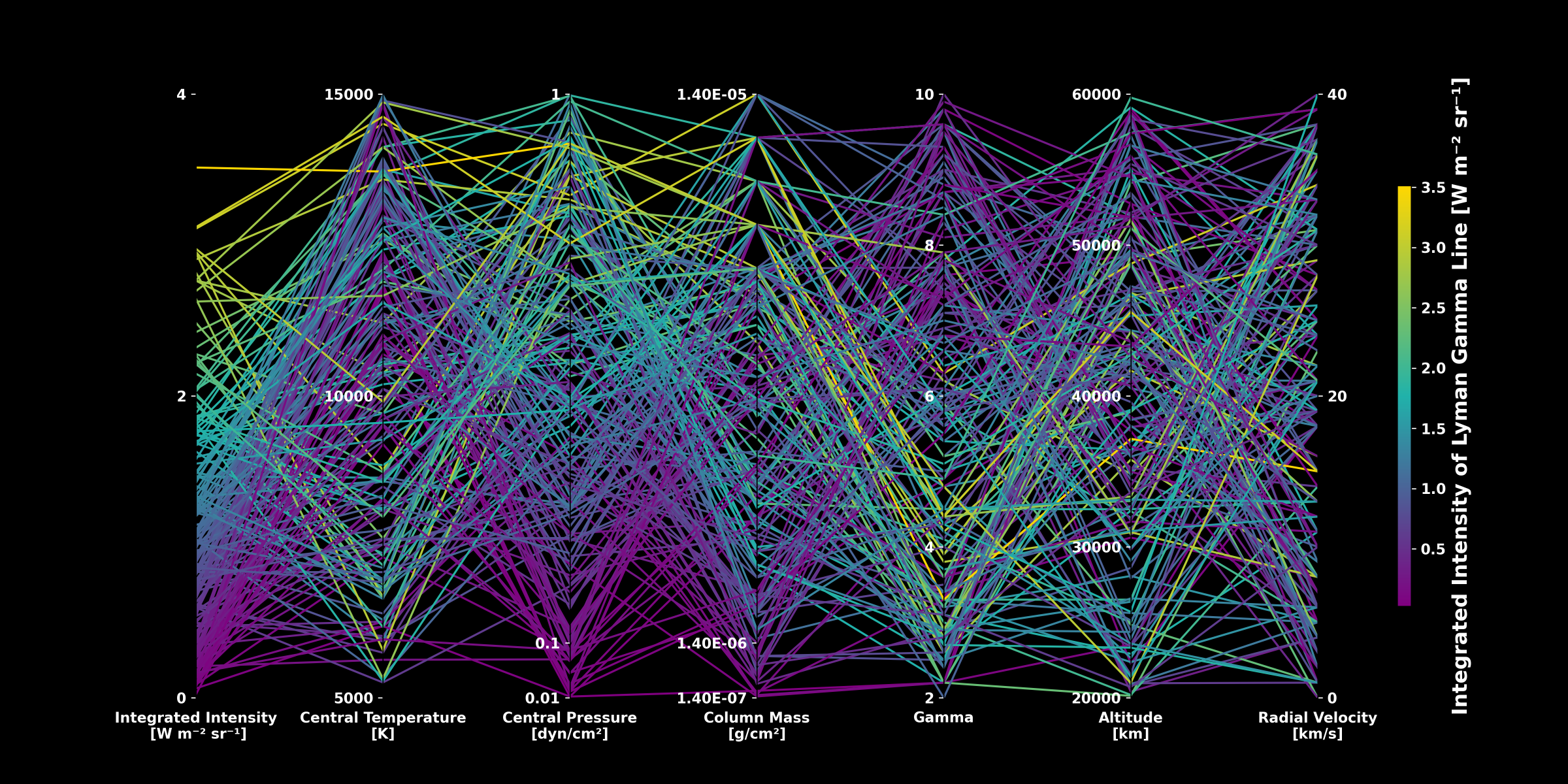}
    \caption{The parallel coordinate plot of 6 parameters with the integrated intensity of the Lyman $\gamma$ line.}
    \label{fig:II973}
\end{figure*}

\begin{figure*}
    \centering
    \includegraphics[width=0.95\textwidth]{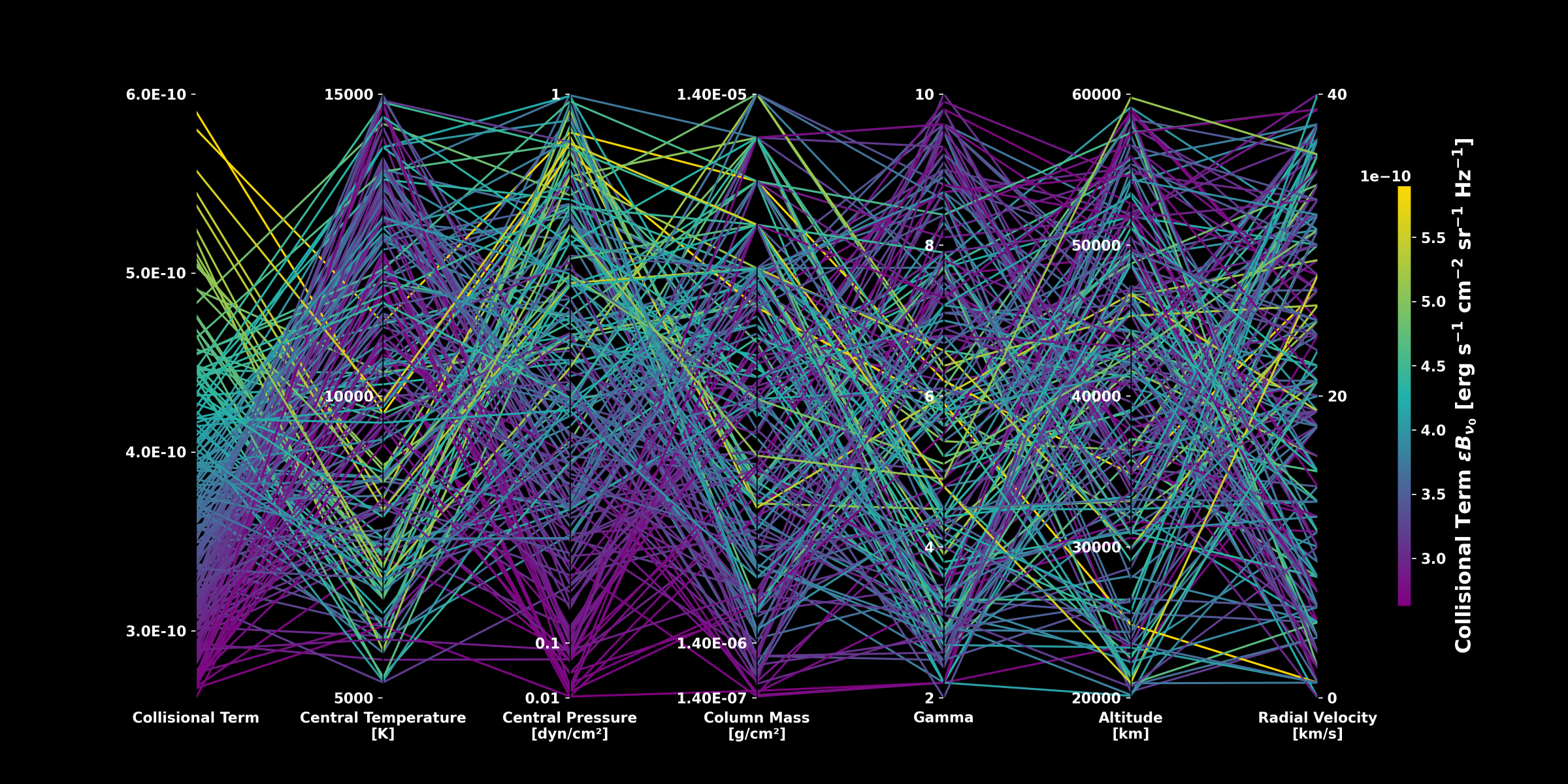}
    \caption{The parallel coordinate plot of 6 parameters with the collisional term $\epsilon B_{\nu_0}$ of the Lyman $\gamma$ line.}
    \label{fig:CT973}
\end{figure*}

\begin{figure*}
    \centering
    \includegraphics[width=0.95\textwidth]{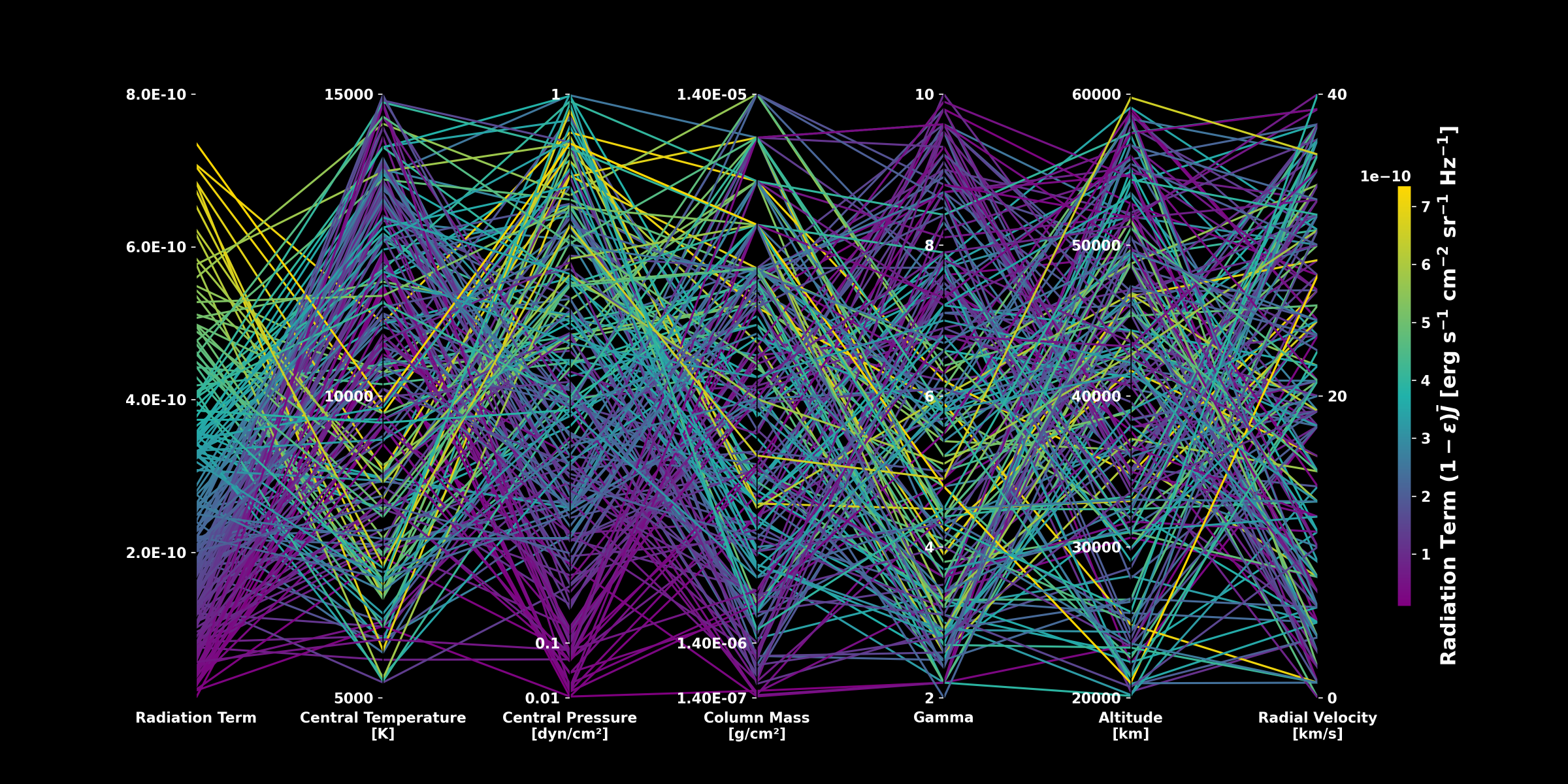}
    \caption{The parallel coordinate plot of 6 parameters with the radiation term $(1 - \epsilon) \bar{J}$ of the Lyman $\gamma$ line.}
    \label{fig:RT973}
\end{figure*}

\begin{figure*}
    \centering
    \includegraphics[width=0.95\textwidth]{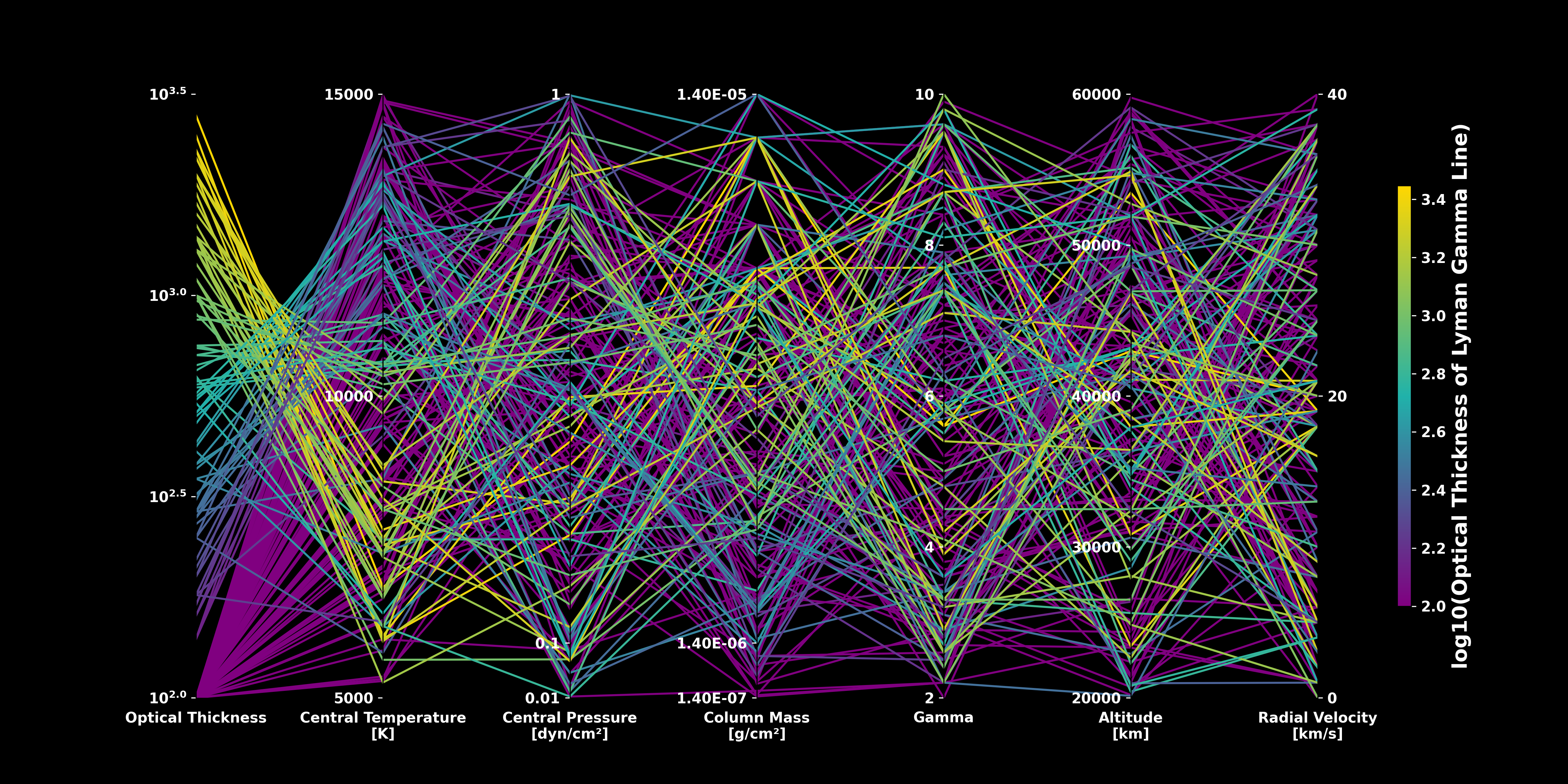}
    \caption{The parallel coordinate plot of 6 parameters with the optical thickness of the Lyman $\gamma$ line.}
    \label{fig:OT973}
\end{figure*}

Figure~\ref{fig:emission_measure} shows the relation between emission measure, temperature, pressure, and slab thickness on the emitted intensity in the H$\alpha$ line.
In a parallel coordinate plot like Figure~\ref{fig:emission_measure}, each line connected from left side to right side represents a model and each connected point on this line represents one parameter of the model. In Figure~\ref{fig:emission_measure}, the temperature, pressure and slab thickness are inputs of the model, and integrated intensity and emission measure are outputs of the model. The color is set by values of integrated intensity. If we see a diversity of colours connected to different value ranges when we focus on one parameter's axis, this means the parameter does not have much effect (such as slab thickness in Figure \ref{fig:emission_measure}); on the other hand, if we see a certain colour concentrated in a certain value range, this means the parameter has some effect (such as emission measure, temperature, and pressure in Figure \ref{fig:emission_measure}). 

Figure~\ref{fig:Heinzel1994para} shows the relations between H$\alpha$ integrated intensity and emission measure, temperature, pressure and slab thickness in a more common representation.
Figure~\ref{fig:Heinzel1994para} shows that H$\alpha$ intensity increases with emission measure and pressure. This conclusion is the same as \cite{heinzel1994theoretical}. The panel showing integrated intensity of the H$\alpha$ line vs. temperature and slab thickness in Figure~\ref{fig:Heinzel1994para} are not included in \cite{heinzel1994theoretical}. This figure shows that H$\alpha$ intensity also increases with temperature. \added{For the models computed in this study, Figure~\ref{fig:Heinzel1994para} shows that the H$\alpha$ integrated intensity vs. slab thickness relationship does not exhibit as clear a trend as the other relationships.}


Figure~\ref{fig:Heinzel1994para} shows the relations between emission measure, temperature, pressure and slab thickness, with the integrated intensity of the H$\alpha$ line. However, a single parallel coordinate plot in Figure \ref{fig:emission_measure}  contains the same information about these relations, and  includes even more information. When the value range on which a certain colour is concentrated is narrow, the effect is more significant. From Figure \ref{fig:emission_measure}, we know emission measure, temperature, and pressure have an effect on the integrated intensity of the H$\alpha$ line. From the concentration of colour, we know the correlation between emission measure and the integrated intensity of the H$\alpha$ line is stronger than the correlation between pressure and the integrated intensity of the H$\alpha$ line, and the correlation between temperature and the integrated intensity of the H$\alpha$ line is even weaker than the correlation between pressure and the integrated intensity of the H$\alpha$ line. We can use correlation coefficients to further confirm this result. After calculation, we know that emission measure, temperature, and pressure have correlation coefficients of 0.98, 0.41, and 0.66 with the integrated intensity of the H$\alpha$ line, which is consistent with our result. It is not necessary to use correlation coefficients when we use parallel coordinate plots. It depends what conclusions the user wants. If the user just want to find out which parameter has stronger effect on a property, it is not necessary to use the correlation coefficient. Actually, the parallel coordinate plot contains more information about how a parameter is related to a property. But if the user want to quantify the result, we can calculate the correlation coefficient.

Additionally, we can not only check one parameter's effect by focusing on the axis, but also track other parameters' values by following the lines. In this way, we can find out the composition of models with high integrated intensity and those with low integrated intensity. 

Parallel coordinate plots complement correlation coefficients by showing how parameter influences vary across their value ranges. In the rest of our analysis, they will serve mainly as a compact way to identify where the collisional term, radiation term, integrated intensity, and optical thickness reach high or low values, and what parameter combinations produce them. This integrated view of multi-parameter behavior cannot be obtained from correlation coefficients alone, making the plot an efficient diagnostic tool for identifying non-linear or range-dependent relationships.

In summary, a parallel coordinate plot has the following advantages:
\begin{itemize}
\item It helps us discover potential relations among different parameters. The parallel coordinate plot effectively illustrates how all data points transition across multiple parameters, thereby revealing structural patterns and potential inter-parameter relationships. We can intuitively draw some conclusions by focusing on a parallel coordinate plot. It is not a quantitative method, but it helps to identify possible correlations among parameters. Quantitative methods can be further used to characterise correlations that are found in a parallel coordinate plot. Especially when the relationship is non-linear, it would be hard to notice the relationship just by quantitative methods like correlation coefficient calculation.
\item It is an efficient way to show parameters' relations. A single parallel coordinate plot contains a large amount of information. Other more traditional figures show similar information but require several 
figures to arrive at the same level of information.
\item We can track specific colors to identify the parameters that are used to construct different models;
\item We can understand the sensitivity of the property that determines the colors to  different ranges of the other parameters shown in the plot.
\end{itemize}

We will introduce these advantages with detailed examples in the following sections.

\subsubsection{Elasticity coefficient}

We calculate the elasticity coefficient to show how the quantities we study (quantities in the first axis and depicted by the colour in our parallel coordinate plots) are sensitive to changes in other parameters. We calculate the elasticity coefficient between two quantities $x$ and $y$ (e.g. central pressure and integrated intensity) by Equation~\ref{ela}. 

\begin{eqnarray}
e_{xy} = \frac{\frac{\Delta x}{\overline{x}}}{\frac{\Delta y}{\overline{y}}}=\frac{\Delta x}{\Delta y} \cdot \frac{\overline{y}}{\overline{x}}
\label{ela}
\end{eqnarray}

$\frac{\Delta x}{\Delta y}$ represents the instantaneous rate of change of $x$ with respect to $y$, which is estimated by the regression coefficient in a simple linear regression model. $\overline{x}$ and $\overline{y}$ are means of $x$ and $y$. A larger elasticity coefficient $e_{xy}$ (the absolute value) means the $y$ parameter is more sensitive to the $x$ parameter's relative change $\frac{\Delta x}{\overline{x}}$. 

We can also take \cite{heinzel1994theoretical} as example. After calculation, we know that emission measure, temperature, and pressure have elasticity coefficients of 5.91, 0.24, and 0.22 with the integrated intensity of the H$\alpha$ line. This suggests that the integrated intensity of the H$\alpha$ line is much more sensitive to emission measure's relative change than other parameters. If we refer to Figure~\ref{fig:Heinzel1994para}, we see that this result is quite accurate. 

In Figure~\ref{fig:emission_measure}, the high-value and low-value ranges of integrated intensity of the H$\alpha$ line are more concentrated at the axis of emission measure and colour in those ranges is more uniform. This characteristic actually suggests the property we study is more sensitive with the change of this parameter, which means that it has a high elasticity coefficient. 

\section{Results}
\label{sec:effect}
\subsection{Effects of model parameters on Lyman line properties}
Using the altitude and incident radiation constraints as described in \added{Section~\ref{sec:obs2} and Section~\ref{Sec:incident}},  the radial velocity constraint from \cite{zhang2026analysis}, and the usual range for the other parameters \added{from }\cite{labrosse2012plasma}, we generate 200 random PCTR models for the prominence as shown in Table \ref{table:parameter_ranges}.

In Figures~\ref{fig:II1026}--\ref{fig:OT973}, we explore the effects of central temperature, central pressure, column mass, gamma, altitude and radial velocity on integrated intensity, collisional term, radiation term and optical thickness. Since there are more obvious effects on a parallel coordinate plot when the correlation coefficient is larger than 0.3, we consider a parameter important in this case. We also use the correlation coefficient and the elasticity coefficient (between the parameter and the property we study) in the Table~\ref{table:beta} and Table~\ref{table:gamma} to quantify the effect. Values not shown in the tables correspond to correlation coefficients smaller than 0.3 and their elasticity coefficients are generally smaller than the lowest values listed in the tables. However, there are exceptions. For example, the elasticity coefficient of central temperature for the radiation term in Lyman $\beta$ and Lyman $\gamma$ lines is very closed to that of column mass (0.39 and 0.42), even though its correlation coefficient is below 0.3. Similarly, for the collisional term in the Lyman $\gamma$ line, altitude and central temperature have elasticity coefficients (-0.15 and 0.13) comparable to that of column mass. For optical thickness, only the elasticity coefficient of temperature exceeds 1, while column mass also shows relatively large values (0.77 for Lyman $\beta$ and 0.62 for Lyman $\gamma$) though its correlation coefficient is below 0.3.


\begin{table}
\centering
\caption{Parameter ranges for different variables}
\label{table:parameter_ranges}
\begin{tabular}{m{3.3cm} m{4cm}}
\hline
Parameter & Range \\
\hline
Central Temperature & $5000~\text{K} \sim 15000~\text{K}$ \\
Surface Temperature & $100000~\text{K}$ \\
Central Pressure & $0.01~\text{dyn}~\text{cm}^{-2} \sim 1.0~\text{dyn}~\text{cm}^{-2}$ \\
Surface Pressure & $0.01~\text{dyn}~\text{cm}^{-2}$ \\
Column Mass & $1.4 \times 10^{-7}~\text{g}~\text{cm}^{-2} \sim 1.4 \times 10^{-5}~\text{g}~\text{cm}^{-2}$ \\
$\gamma$ & $2 \sim 10$ \\
Altitude & $20000~\text{km} \sim 60000~\text{km}$ \\
Radial Velocity & $0~\text{km}~\text{s}^{-1} \sim 40~\text{km}~\text{s}^{-1}$ \\
Microturbulent Velocity & $5~\text{km}~\text{s}^{-1}$ \\
\hline
\end{tabular}
\end{table}

\begin{table}
\centering
\caption{Correlation and elasticity coefficients for Lyman $\beta$ line properties}
\label{tab:lyb_combined}
\begin{tabular}{m{2.5cm}ccccc}
\hline
\multirow{2}{*}{Property} & \multirow{2}{*}{Key Parameters} & \multicolumn{2}{c}{Coefficients} \\
\cline{3-4}
 & & Correlation & Elasticity \\
\hline
\multirow{3}{*}{Integrated Intensity} 
 & Central pressure & 0.70 & 0.90 \\
 & Column mass & 0.49 & 0.61 \\
 & Gamma & -0.32 & -0.56 \\
\hline
\multirow{2}{*}{Collisional Term $\epsilon B_{\nu_0}$} 
 & Central pressure & 0.66 & 0.35 \\
 & Column Mass & 0.37 & 0.19 \\
 \hline
\multirow{2}{*}{Radiation Term $(1 - \epsilon) \bar{J}$} 
 & Central pressure & 0.72 & 0.86 \\
 & Column Mass & 0.34 & 0.39 \\
 & Gamma & -0.34 & -0.54 \\
\hline
\multirow{2}{*}{Optical Thickness} 
 & Central temperature & -0.66 & -2.13 \\
 & Column mass & 0.51 & 0.76 \\
\hline
\end{tabular}
\begin{flushleft}
Note: The parameters that don't appear in the table correspond to correlation coefficients less than 0.3.
\end{flushleft}
\label{table:beta}
\end{table}

\begin{table}
\centering
\caption{Correlation and elasticity coefficients for Lyman $\gamma$ line properties}
\label{tab:lyg_combined}
\begin{tabular}{m{2.5cm}ccccc}
\hline
\multirow{2}{*}{Property} & \multirow{2}{*}{Key Parameters} & \multicolumn{2}{c}{Coefficients} \\
\cline{3-4}
 & & Correlation & Elasticity \\
\hline
\multirow{3}{*}{Integrated Intensity} 
 & Central pressure & 0.69 & 0.89 \\
 & Column mass & 0.45 & 0.56 \\
 & Gamma & -0.37 & -0.65 \\
\hline
\multirow{2}{*}{Collisional Term $\epsilon B_{\nu_0}$} 
 & Central pressure & 0.69 & 0.25 \\
 & Column Mass & 0.37 & 0.13 \\
 \hline
\multirow{2}{*}{Radiation Term $(1 - \epsilon) \bar{J}$} 
 & Central pressure & 0.69 & 0.86 \\
 & Column Mass & 0.35 & 0.42 \\
 & Gamma & -0.34 & -0.56 \\
\hline
Optical Thickness 
 & Central temperature & -0.37 & -2.23 \\
\hline
\end{tabular}
\label{table:gamma}
\begin{flushleft}
Note: The parameters that don't appear in the table correspond to correlation coefficients less than 0.3.
\end{flushleft}
\end{table}

From Figures~\ref{fig:II1026}--\ref{fig:OT973}, the dependencies between the line properties and the model parameters can be summarized as follows. For both Lyman $\beta$ and Lyman $\gamma$, a higher central pressure usually corresponds to a higher integrated intensity, while larger values of column mass also tend to increase the intensity. In contrast, when $\gamma$ increases, the integrated intensity decreases, indicating that a stronger temperature gradient suppresses the total emission. 

For the collisional term $\epsilon B_{\nu_0}$, higher central pressure or larger column mass is generally associated with a larger collisional term. For the radiation term $(1 - \epsilon) \bar{J}$, higher central pressure or larger column mass is also generally associated with a larger radiation term, whereas larger $\gamma$ values lead to smaller values of the radiation term. 


For the optical thickness, an increase in central temperature tends to reduce optical thickness in both lines, meaning that hotter core conditions lead to smaller optical thickness. At the same time, larger column mass values correspond to higher optical thickness, reflecting the role of plasma content in enhancing absorption. This effect is strong in Lyman $\beta$ line but not in Lyman $\gamma$ line.

\subsection{The Lyman $\beta$ line}

Figures~\ref{fig:II1026}--\ref{fig:OT1026} show the relationships between various parameters and the Lyman $\beta$ line properties. In Figure~\ref{fig:II1026}, we can see that in the axis of central pressure, column mass and gamma, the colour seems to be more uniform in different parts. This usually suggests a relatively high correlation coefficient (larger than 0.3). Also, the axis of central pressure shows less diversity of colour in different parts of the axis, which suggests a higher correlation coefficient. We can refer to Table~\ref{tab:lyb_combined}. The correlation coefficient between central pressure and integrated intensity of Lyman $\beta$ is 0.70. This is higher than column mass and gamma. In Figure~\ref{fig:CT1026} and Figure~\ref{fig:RT1026}, we can see that the collisional term $\epsilon B_{\nu_0}$, and the radiation term $(1 - \epsilon) \bar{J}$ have a similar order of magnitude and radiation term is about $\sim$ 2 times of collisional term. This suggests that the collisional and radiative contributions are equally important for the formation of the Lyman $\beta$ line.


The elasticity coefficients for the Lyman $\beta$ line properties are presented in Table~\ref{tab:lyb_combined}. For parameters with large values of the elasticity coefficient in Table~\ref{tab:lyb_combined}, we can refer to Figure~\ref{fig:II1026}, Figure~\ref{fig:RT1026}, and Figure~\ref{fig:OT1026}. The high-value and low-value ranges of the quantities we study are more concentrated and their colour in those ranges is more uniform. In Figure~\ref{fig:OT1026}, the high-value range of the optical thickness (larger than 4000) is well concentrated below the 10000~K on the axis of the central temperature, while the values of the optical thickness are all below 4000 above 10000~K. Also, in Figure~\ref{fig:RT1026}, we can see that the high-value range of the radiation term $(1 - \epsilon) \bar{J}$ (larger than $2\times 10^{-9}$) is well concentrated on the upper side of the axis of the central pressure. Therefore, the elasticity coefficient is also a useful tool to describe the characteristics of certain parameters. When an elasticity coefficient for two parameters is high, this indicates that it is possible to use one parameter to constrain the other (we will discuss this further in Section~\ref{sec:refine}). 

The correlation coefficients and elasticity coefficients of the key parameters for the properties of the Lyman $\beta$ line are summarized in Table~\ref{tab:lyb_combined}. From Table~\ref{tab:lyb_combined}, we can see that there are no direct relations between the correlation coefficients and the elasticity coefficients. The elasticity coefficients can be high or low with a similar correlation coefficient.

\subsection{The Lyman $\gamma$ line}

The correlation coefficients and elasticity coefficients of the key parameters for the properties of the Lyman $\gamma$ line are summarized in Table~\ref{tab:lyg_combined}. The results are similar to those obtained from the analysis of the Lyman $\beta$ line. 

In Figure~\ref{fig:II973}, we can see that in the axis of central pressure, column mass and gamma, the colour seems to be more uniform in different parts, which suggests a relatively high correlation coefficient (larger than 0.3). In Figures~\ref{fig:CT973} and \ref{fig:RT973}, we can see that the collisional term $\epsilon B_{\nu_0}$ and radiation term $(1 - \epsilon) \bar{J}$ have a similar distribution of values. This suggests that collisional contribution and radiative contribution are equally important for the formation of the Lyman $\gamma$ line. In Figures~\ref{fig:OT973}, we can see that all yellow lines (optical thickness larger than 1000) distribute below  10000~K on the axis of central temperature and the distribution seems to be random on other axis. This suggests that temperature is the only key parameter that is related to optical thickness of the Lyman $\gamma$ line.

\subsection{Constraining parameters using SPICE data}\label{sec:refine}

\begin{figure}
    \centering

    \begin{minipage}{0.48\textwidth}
        \centering
        \includegraphics[width=\linewidth]{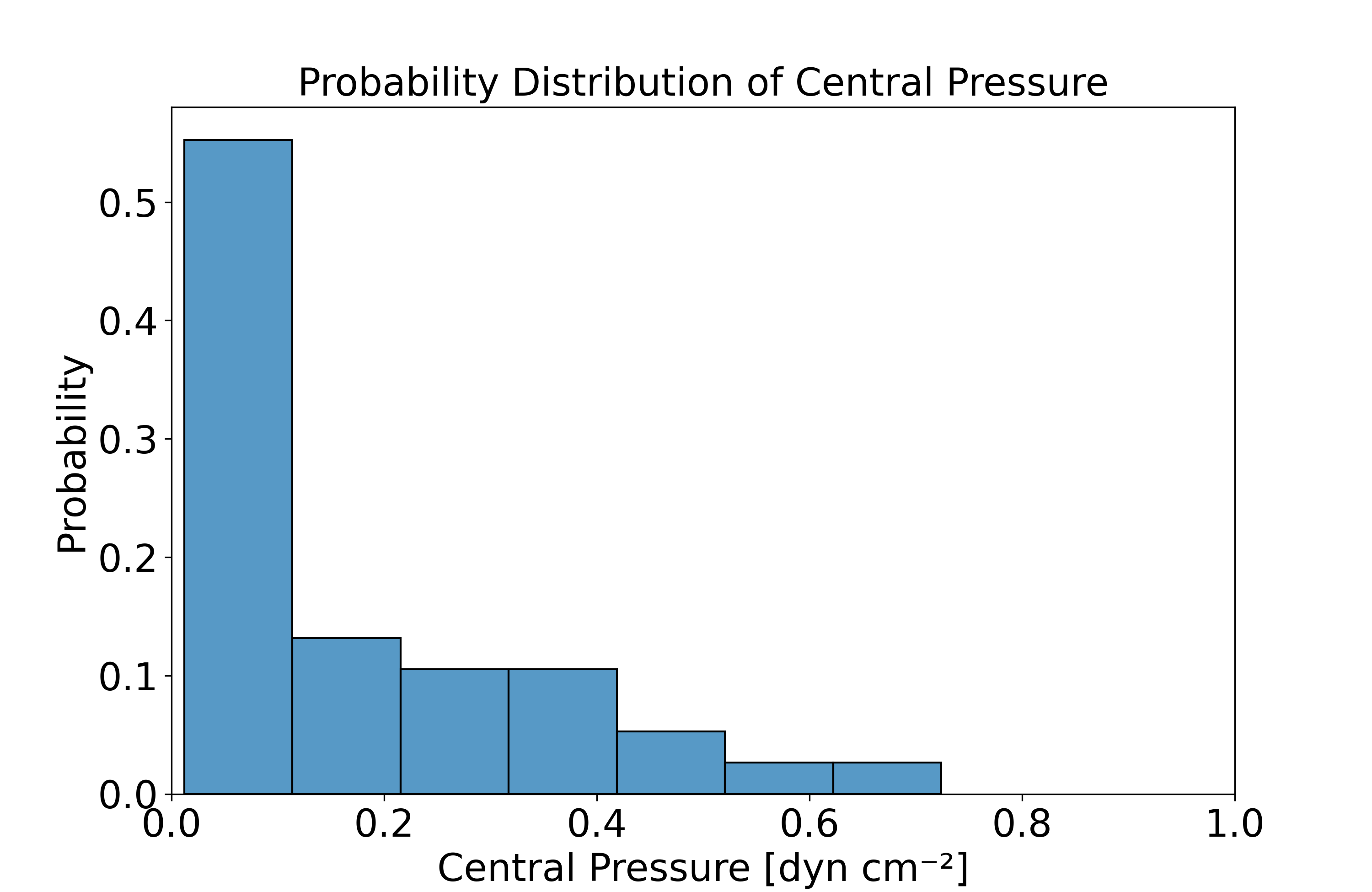}
        \caption{The probability distribution of the central pressure refined by observation.}
        \label{fig:pdf_pre}
    \end{minipage}
    \hfill
    \begin{minipage}{0.48\textwidth}
        \centering
        \includegraphics[width=\linewidth]{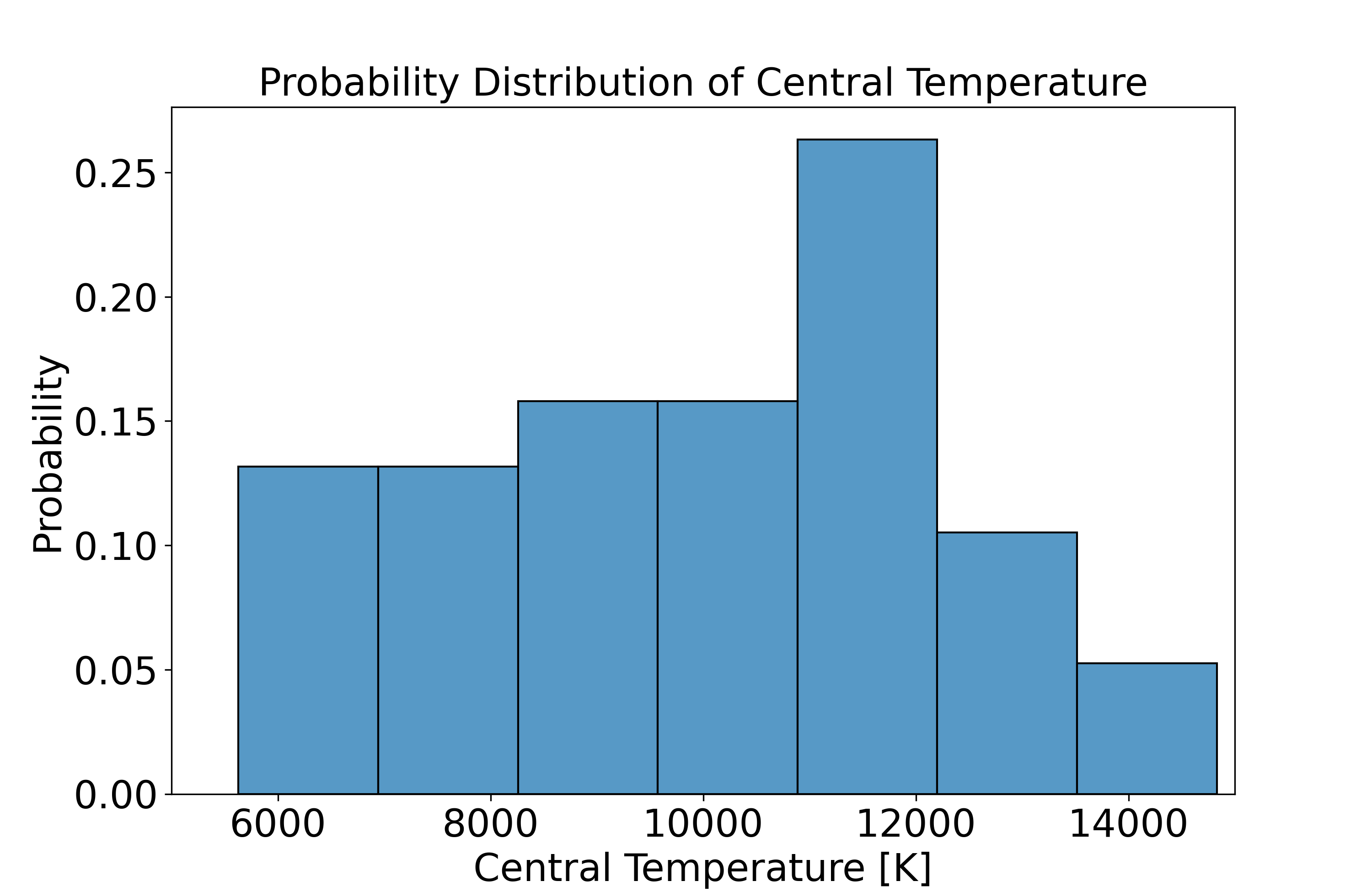}
        \caption{The probability distribution of the central temperature refined by observation.}
        \label{fig:pdf_tem}
    \end{minipage}

\end{figure}

\begin{figure}
    \centering

    \begin{minipage}{0.48\textwidth}
        \centering
        \includegraphics[width=\linewidth]{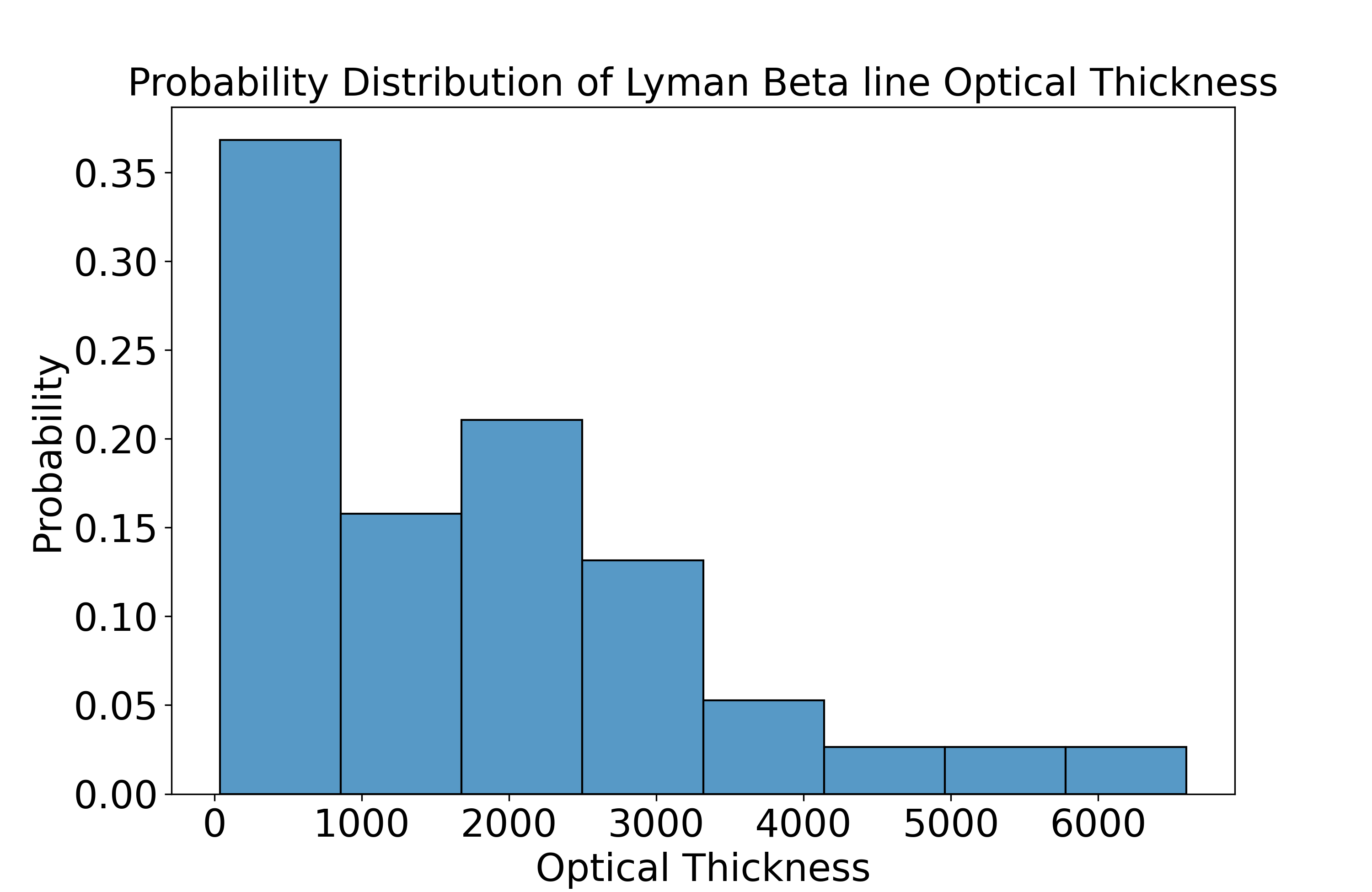}
        \caption{\added{The probability distribution of the Lyman $\beta$ line optical thickness refined by observation.}}
        \label{fig:pdf_otb}
    \end{minipage}
        \hfill
    \begin{minipage}{0.48\textwidth}
        \centering
        \includegraphics[width=\linewidth]{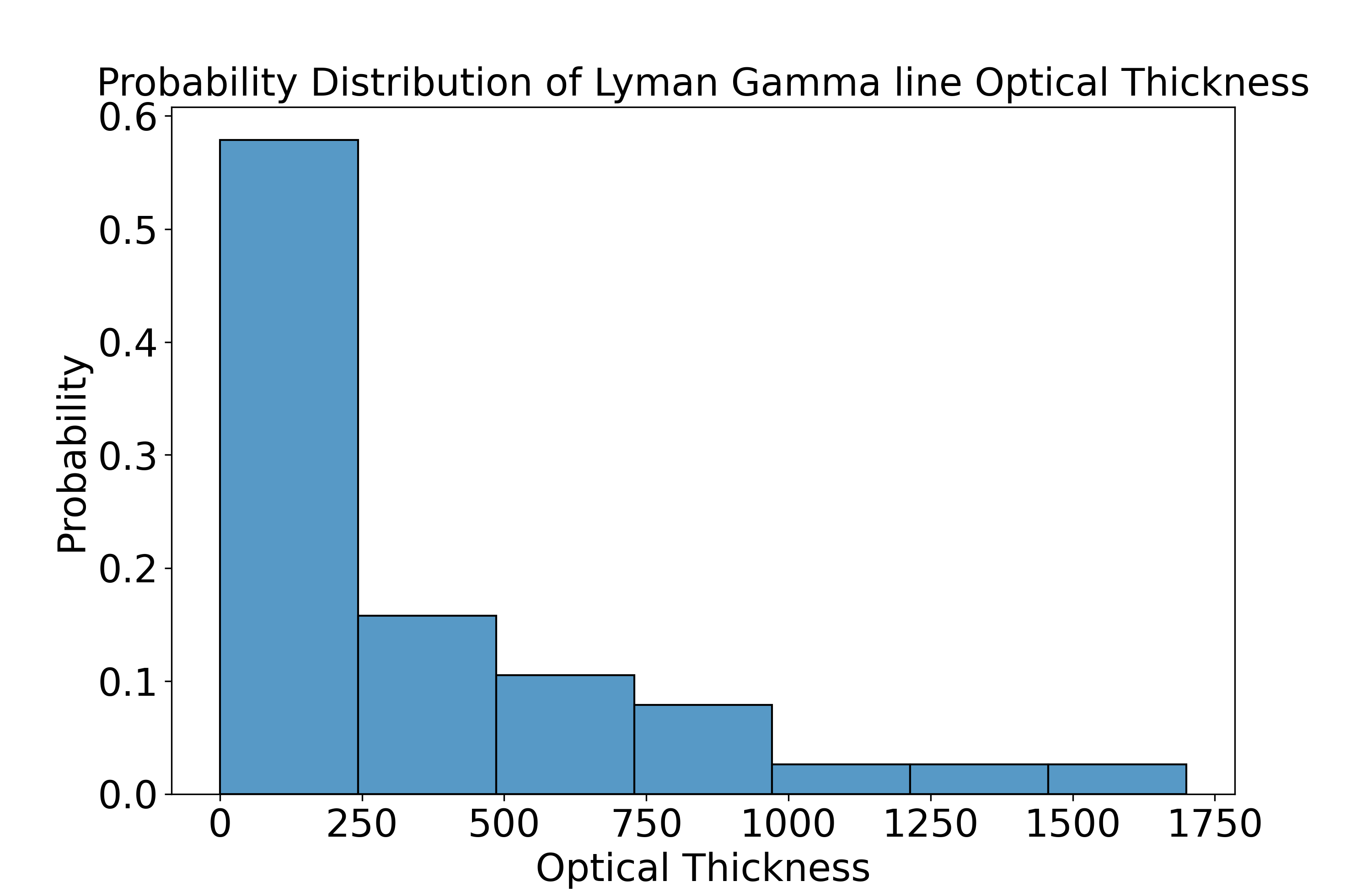}
        \caption{\added{The probability distribution of the Lyman $\gamma$ line optical thickness refined by observation.}}
        \label{fig:pdf_otg}
    \end{minipage}

\end{figure}
We have just determined that the important parameters for the integrated intensity of the Lyman $\beta$ and the Lyman $\gamma$ lines are central pressure, column mass, and gamma. Therefore, we can further constrain these parameters through the Lyman $\beta$ and the Lyman $\gamma$ line intensities measured for the prominence. In Figure~\ref{fig:beta_gamma}, the red rectangle is the prominence region we study. The mean integrated intensity of  Lyman $\beta$ in this region is 0.46~W m$^{-2}$ sr$^{-1}$, and the standard deviation is 0.21~W m$^{-2}$ sr$^{-1}$. The mean integrated intensity of  Lyman $\gamma$ in this region is 0.16~W m$^{-2}$ sr$^{-1}$, and the standard deviation is 0.08~W m$^{-2}$ sr$^{-1}$. 

When we apply a restriction to filter the 200 random models to make the Lyman $\beta$ and  Lyman $\gamma$ integrated intensities within three standard deviations of the mean, we obtain 38 models which meet the requirement. We find that 36 of them have a central pressure below 0.48~$\text{dyn}\,\text{cm}^{-2}$. However, we find that the column mass and gamma of these models seem to have a nearly equal possibility of being distributed in each value within the range after restricting the integrated intensity of the Lyman $\beta$ and Lyman $\gamma$ lines. This might be because the correlation coefficient and the elasticity coefficient between the integrated intensity and the column mass, and also that between the integrated intensity and  gamma, are not as large as those between the integrated intensity and the central pressure. We do not refine all parameters but only parameters with clear biased distributions that show systematic tendencies for data to cluster in certain regions rather than being balanced or symmetric across the entire range.. 

Figure~\ref{fig:pdf_pre} shows the probability distribution of the central pressure after restricting  the models by considering the observed integrated intensities. The distribution is derived from the relative frequencies of central pressure values, calculated by counting how many models fall within each pressure interval and normalizing by the total number of accepted models. There are no models for values above 0.7~$\text{dyn}\,\text{cm}^{-2}$, and models with lower pressures are more frequent. Thus, central pressure is a good example of a parameter we can constrain.

Figure~\ref{fig:pdf_tem} shows the probability distribution of the central temperature after constraining the models by considering the observed integrated intensities. The possibility of values lying around the middle of the range of temperatures are slightly higher than other ranges of values. Nevertheless, other temperature values are likely. Therefore, we would not constrain the parameters with a distribution like that in Figure~\ref{fig:pdf_tem}. 

Since 95~\% of the models that meet the restriction requirement of both lines have a central pressure below 0.48~$\text{dyn}\,\text{cm}^{-2}$, we can judge that the central pressure is below this value. For other parameters,  we have to find other lines which have stronger relations with those parameters to refine them in the future. 

\added{As shown in Figure~\ref{fig:pdf_otb} and Figure~\ref{fig:pdf_otg}, we can also generate the probability distribution of Lyman $\beta$ and Lyman $\gamma$ optical thicknesses after constraining the models by considering the observed integrated intensities. It also shows a biased distribution. Since 95~\% of the models that meet the restriction requirement have Lyman $\beta$ line optical thickness below 4800 and Lyman $\gamma$ line optical thickness below 1000, we can constrain the Lyman $\beta$ line optical thickness and the Lyman $\gamma$ line optical thickness by these values. 
The optical thickness is an output parameter of the PRODOP code, and these figures give us an indication of what the expected optical thickness is in these two lines. This result demonstrates that extremely optically thick models for Lyman $\beta$ and Lyman $\gamma$ lines are strongly excluded by the observations.}

In summary, this method using non-LTE modelling results to constrain plasma parameters from observations has the potential to be extended to more lines. This will give better constraints on  prominence parameters.

\section{Discussion}\label{sec:discussion}
We have investigated the most important parameters in the 2023 April 15th event for the formation of the Lyman $\beta$ line and the Lyman $\gamma$ line. The most important parameters are central temperature, central pressure, column mass, and the temperature gradient in the prominence-to-corona transition region, which we can see from Figures~\ref{fig:II1026}--\ref{fig:OT973}. The correlation coefficient describes how strong the relations of these parameters are with the quantities we study and the elasticity coefficient describes how sensitive the quantities are to the relative changes of these parameters. The parameters' effects among the Lyman $\beta$ line and the Lyman $\gamma$ line are quite similar. We can interpret how these influences take place.

The temperature affects the distribution of different energy levels' populations which then affects optical thickness. 

The pressure affects the density and level populations of hydrogen atoms and the density of electrons, and therefore affects the collisional contribution and the integrated intensity. The density should also affect the optical thickness, but in the context of this event, we can see that the effect is not so strong in Tables~\ref{table:beta} and \ref{table:gamma}.

The column mass affects the integrated intensity and the collisional contribution. The column mass affects the radiation term $(1 - \epsilon) \bar{J}$ in the expression of the source function (Eq. \ref{Eq: source}).  When we change the column mass, the density of particles does not change much. The total intensity of the radiation field in a particular direction would have a positive correlation with column mass (larger volume). Therefore, the change in the column mass can affect the collisional contribution. Through source function, the column mass can affect the integrated intensity indirectly.

The value of gamma (corresponding to the steepness of the temperature gradient in the PCTR) affects the integrated intensity and the collisional contribution. When gamma is high, it means the temperature changes more quickly in the PCTR, which means the high-temperature part of prominence is less extended. Therefore, it impacts the temperature profile inside the prominence, and thus affects the integrated intensity and the collisional contribution to the source function.

\section{Conclusions}\label{sec:conclusion}
By comparing observed Lyman line intensities with the results of non-LTE modelling, this study extends our analysis presented in \cite{zhang2026analysis} of the first dedicated observation of an off-limb prominence by the Spectral Imaging of the Coronal Environment (SPICE) which took place on April 15, 2023. The prominence was observed a few hours before its eruption. We used a 1D non-LTE radiative transfer model (PRODOP) to investigate how different physical parameters affect the formation of the Lyman $\beta$ and Lyman $\gamma$ lines of hydrogen. The incident radiation for these lines was constrained and re-scaled using the SPICE full-disk mosaic observation from November 13, 2023, providing a more realistic boundary condition for the radiative transfer calculations. 

We introduced parallel coordinate plots as an effective tool for visualizing multi-parameter relationships with different properties. To demonstrate the relevance of this toolkit, we  confirmed the well-known correlation between H$\alpha$ intensity and emission measure, but also revealed an increase of H$\alpha$ intensity with temperature which is not discussed in previous studies. We also use the parallel coordinate plot and the elasticity coefficient to further compare relations among different parameters. The emission measure has the strongest relation with H$\alpha$ intensity, and the H$\alpha$ intensity is much more sensitive to the emission measure's relative change than other parameters.

With respect to the SPICE prominence observations discussed in \cite{zhang2026analysis} and summarised here, our results further show that central temperature, central pressure, column mass, and PCTR temperature gradient are important parameters for the formation of the Lyman $\beta$ and Lyman $\gamma$ lines. In addition, one can also use the relations between these parameters and the observed integrated intensities to refine the parameter ranges. As a result, we find that the central pressure of this prominence should be below 0.48~$\text{dyn}/\text{cm}^2$\added{, Lyman $\beta$ line optical thickness should be below 4800 and Lyman $\gamma$ line optical thickness should be below 1000}.

In a future work, we plan to extend this analysis to additional spectral lines that the PRODOP code can calculate. Combining these with data from other instruments, e.g. H$\alpha$ from ground-based observatories, EUI on Solar Orbiter, and IRIS observations of Mg~II lines \citep{xue2025multiwavelength}, will allow us to compare data with non-LTE models and perform a multi-line analysis that can better constrain a wider range of prominence parameters. This approach will enhance our ability to infer the thermodynamic and dynamic properties of solar prominences, ultimately improving our understanding of evolution of plasma parameters before and during eruption. Since multi-line non-LTE diagnostics have potentials to constrain the temperature, density, ionisation degree, and velocity of the prominence in the pre-eruption phase, they can also provide indirect constraints on the physical processes responsible for triggering eruptions.

\section*{Acknowledgments}
We would like to thank the teams involved in the development and operation of the SPICE instrument and data release. We are also grateful to the SunPy team for providing the tools used in our data analysis. The authors thank Lyndsay Fletcher for her comments on this paper. We also thank the referee for the efficient review process. YZ is supported by the China Scholarship Council (No. 202206120056). NL acknowledges support from UK Research and Innovation's Science and Technology Facilities Council under grant award numbers ST/T000422/1 and ST/X000990/1. TAK was supported by Solar Orbiter/SPICE funding to NASA Goddard Space Flight Center.

Solar Orbiter is a space mission of international collaboration between ESA and NASA, operated by ESA. The development of SPICE has been funded by ESA member states and ESA. It was built and is operated by a multi-national consortium of research institutes supported by their respective funding agencies: STFC RAL (UKSA, hardware lead), IAS (CNES, operations lead), GSFC (NASA), MPS (DLR), PMOD/WRC (Swiss Space Office), SwRI (NASA), UiO (Norwegian Space Agency).

\section*{Data Availability}

This research used version 5.0.0 of the SunPy open source software package. We obtained SPICE data through the ESA Solar Orbiter Archive. We have used SPICE Data Release 5.0: (DOI: https://doi.org/10.48326/idoc.medoc.spice.5.0).



\bibliographystyle{mnras}
\bibliography{paper2} 

@ARTICLE{muller2020solar,
       author = {{M{\"u}ller}, D. and {St. Cyr}, O.~C. and {Zouganelis}, I. and {Gilbert}, H.~R. and {Marsden}, R. and {Nieves-Chinchilla}, T. and {Antonucci}, E. and {Auch{\`e}re}, F. and {Berghmans}, D. and {Horbury}, T.~S. and {Howard}, R.~A. and {Krucker}, S. and {Maksimovic}, M. and {Owen}, C.~J. and {Rochus}, P. and {Rodriguez-Pacheco}, J. and {Romoli}, M. and {Solanki}, S.~K. and {Bruno}, R. and {Carlsson}, M. and {Fludra}, A. and {Harra}, L. and {Hassler}, D.~M. and {Livi}, S. and {Louarn}, P. and {Peter}, H. and {Sch{\"u}hle}, U. and {Teriaca}, L. and {del Toro Iniesta}, J.~C. and {Wimmer-Schweingruber}, R.~F. and {Marsch}, E. and {Velli}, M. and {De Groof}, A. and {Walsh}, A. and {Williams}, D.},
        title = "{The Solar Orbiter mission. Science overview}",
      journal = {\aap},
         year = 2020,
        month = oct,
       volume = {642},
          eid = {A1},
        pages = {A1},
          doi = {10.1051/0004-6361/202038467},
archivePrefix = {arXiv},
       eprint = {2009.00861},
 primaryClass = {astro-ph.SR},
       adsurl = {https://ui.adsabs.harvard.edu/abs/2020A&A...642A...1M}
}

@ARTICLE{labrosse2012plasma,
       author = {{Labrosse}, N. and {McGlinchey}, K.},
        title = "{Plasma diagnostic in eruptive prominences from SDO/AIA observations at 304 {\r{A}}}",
      journal = {\aap},
         year = 2012,
        month = jan,
       volume = {537},
          eid = {A100},
        pages = {A100},
          doi = {10.1051/0004-6361/201117801},
archivePrefix = {arXiv},
       eprint = {1111.4847},
 primaryClass = {astro-ph.SR},
       adsurl = {https://ui.adsabs.harvard.edu/abs/2012A&A...537A.100L}
}

@ARTICLE{rochus2020solar,
       author = {{Rochus}, P. and {Auch{\`e}re}, F. and {Berghmans}, D. and {Harra}, L. and {Schmutz}, W. and {Sch{\"u}hle}, U. and {Addison}, P. and {Appourchaux}, T. and {Aznar Cuadrado}, R. and {Baker}, D. and et al.},
        title = "{The solar orbiter EUI instrument: the extreme ultraviolet imager}",
      journal = {\aap},
         year = 2020,
        month = oct,
       volume = {642},
          eid = {A8},
        pages = {A8},
          doi = {10.1051/0004-6361/201936663},
       adsurl = {}
}

@ARTICLE{anderson2020solar,
       author = {{Anderson}, M. and {Appourchaux}, T. and {Auch{\`e}re}, F. and {Aznar Cuadrado}, R. and {Barbay}, J. and {Baudin}, F. and {Beardsley}, S. and {Bocchialini}, K. and {Borgo}, B. and {Bruzzi}, D. and et al.},
        title = "{The Solar Orbiter SPICE instrument - An extreme UV imaging spectrometer}",
      journal = {\aap},
         year = 2020,
        month = oct,
       volume = {642},
          eid = {A14},
        pages = {A14},
          doi = {10.1051/0004-6361/201936697},
       adsurl = {}
}

@ARTICLE{labrosse2007effect,
       author = {{Labrosse}, N. and {Gouttebroze}, P. and {Vial}, J.-C.},
        title = "{Effect of motions in prominences on the helium resonance lines in the extreme ultraviolet}",
      journal = {\aap},
         year = 2007,
        month = feb,
       volume = {463},
        pages = {1171--1179},
          doi = {10.1051/0004-6361:20066419},
       adsurl = {}
}

@ARTICLE{labrosse2008diagnostics,
       author = {{Labrosse}, N. and {Vial}, J.-C. and {Gouttebroze}, P.},
        title = "{Diagnostics of active and eruptive prominences through hydrogen and helium lines modelling}",
      journal = {Annales Geophysicae},
         year = 2008,
        month = oct,
       volume = {26},
        pages = {2961--2965},
          doi = {10.5194/angeo-26-2961-2008},
       adsurl = {}
}

@ARTICLE{zhang2019prominences,
       author = {{Zhang}, P.},
        title = "{Prominences and their eruptions as observed with the IRIS mission and ancillary instruments}",
      journal = {Ph.D. Thesis},
         year = 2019,
       adsurl = {}
}

@ARTICLE{anzer1999energy,
       author = {{Anzer}, U. and {Heinzel}, P.},
        title = "{The energy balance in solar prominences}",
      journal = {\aap},
         year = 1999,
        month = sep,
       volume = {349},
        pages = {974--984},
       adsurl = {}
}

@ARTICLE{heinzel2001soho,
       author = {{Heinzel}, P. and {Schmieder}, B. and {Vial}, J.-C. and {Kotr{\v{c}}}, P.},
        title = "{SOHO/SUMER observations and analysis of the hydrogen Lyman spectrum in solar prominences}",
      journal = {\aap},
         year = 2001,
        month = may,
       volume = {370},
        pages = {281--297},
          doi = {10.1051/0004-6361:20010219},
       adsurl = {}
}

@ARTICLE{warren1998high,
       author = {{Warren}, H.~P. and {Mariska}, J.~T. and {Wilhelm}, K.},
        title = "{High-resolution observations of the solar hydrogen Lyman lines in the quiet sun with the SUMER instrument on SOHO}",
      journal = {\apjs},
         year = 1998,
        month = sep,
       volume = {119},
        pages = {105},
          doi = {10.1086/313155},
       adsurl = {}
}

@article{edsall2003parallel,
  title={The parallel coordinate plot in action: design and use for geographic visualization},
  author={Edsall, Robert M},
  journal={Computational Statistics \& Data Analysis},
  volume={43},
  number={4},
  pages={605--619},
  year={2003},
  publisher={Elsevier}
}

@ARTICLE{labrosse2010,
       author = {{Labrosse}, N. and {Heinzel}, P. and {Vial}, J. -C. and {Kucera}, T. and {Parenti}, S. and {Gun{\'a}r}, S. and {Schmieder}, B. and {Kilper}, G.},
        title = "{Physics of Solar Prominences: I{\textemdash}Spectral Diagnostics and Non-LTE Modelling}",
      journal = {\ssr},
         year = 2010,
        month = apr,
       volume = {151},
       number = {4},
        pages = {243-332},
          doi = {10.1007/s11214-010-9630-6},
archivePrefix = {arXiv},
       eprint = {1001.1620},
 primaryClass = {astro-ph.SR},
       adsurl = {https://ui.adsabs.harvard.edu/abs/2010SSRv..151..243L}
}

@PROCEEDINGS{Vial2015,
        title = "{Solar Prominences}",
    booktitle = {Solar Prominences},
         year = 2015,
       editor = {{Vial}, Jean-Claude and {Engvold}, Oddbj{\o}rn},
       series = {Astrophysics and Space Science Library},
       volume = {415},
        month = jan,
          doi = {10.1007/978-3-319-10416-4},
       adsurl = {https://ui.adsabs.harvard.edu/abs/2015ASSL..415.....V}
}

@article{heinzel1994theoretical,
  title={Theoretical correlations between prominence plasma parameters and the emitted radiation},
  author={Heinzel, P and Gouttebroze, P and Vial, J-C},
  journal={Astronomy and Astrophysics (ISSN 0004-6361), vol. 292, no. 2, p. 656-668},
  volume={292},
  pages={656--668},
  year={1994}
}

@article{gouttebroze1993hydrogen,
  title={The hydrogen spectrum of model prominences},
  author={Gouttebroze, P and Heinzel, P and Vial, JC},
  journal={Astronomy and Astrophysics Supplement Series (ISSN 0365-0138), vol. 99, no. 3, p. 513-543.},
  volume={99},
  pages={513--543},
  year={1993}
}

@article{ge2009exploring,
  title={Exploring uncertainty in remotely sensed data with parallel coordinate plots},
  author={Ge, Yong and Li, Sanping and Lakhan, V Chris and Lucieer, Arko},
  journal={International Journal of Applied Earth Observation and Geoinformation},
  volume={11},
  number={6},
  pages={413--422},
  year={2009},
  publisher={Elsevier}
}

@article{gunar2020quiet,
  title={Quiet-Sun hydrogen Lyman-$\alpha$ line profile derived from SOHO/SUMER solar-disk observations},
  author={Gun{\'a}r, S and Schwartz, P and Koza, J and Heinzel, P},
  journal={Astronomy \& Astrophysics},
  volume={644},
  pages={A109},
  year={2020},
  publisher={EDP Sciences}
}

@article{gouttebroze2000ready,
  title={A ready-made code for the computation of prominence NLTE models},
  author={Gouttebroze, P and Labrosse, N},
  journal={Solar Physics},
  volume={196},
  number={2},
  pages={349--355},
  year={2000},
  publisher={Springer}
}

@ARTICLE{2020A&A...642A..14S,
       author = {{SPICE Consortium} and {Anderson}, M. and {Appourchaux}, T. and {Auch{\`e}re}, F. and {Aznar Cuadrado}, R. and {Barbay}, J. and {Baudin}, F. and {Beardsley}, S. and {Bocchialini}, K. and {Borgo}, B. and {Bruzzi}, D. and {Buchlin}, E. and {Burton}, G. and {B{\"u}chel}, V. and {Caldwell}, M. and {Caminade}, S. and {Carlsson}, M. and {Curdt}, W. and {Davenne}, J. and {Davila}, J. and {Deforest}, C.~E. and {Del Zanna}, G. and {Drummond}, D. and {Dubau}, J. and {Dumesnil}, C. and {Dunn}, G. and {Eccleston}, P. and {Fludra}, A. and {Fredvik}, T. and {Gabriel}, A. and {Giunta}, A. and {Gottwald}, A. and {Griffin}, D. and {Grundy}, T. and {Guest}, S. and {Gyo}, M. and {Haberreiter}, M. and {Hansteen}, V. and {Harrison}, R. and {Hassler}, D.~M. and {Haugan}, S.~V.~H. and {Howe}, C. and {Janvier}, M. and {Klein}, R. and {Koller}, S. and {Kucera}, T.~A. and {Kouliche}, D. and {Marsch}, E. and {Marshall}, A. and {Marshall}, G. and {Matthews}, S.~A. and {McQuirk}, C. and {Meining}, S. and {Mercier}, C. and {Morris}, N. and {Morse}, T. and {Munro}, G. and {Parenti}, S. and {Pastor-Santos}, C. and {Peter}, H. and {Pfiffner}, D. and {Phelan}, P. and {Philippon}, A. and {Richards}, A. and {Rogers}, K. and {Sawyer}, C. and {Schlatter}, P. and {Schmutz}, W. and {Sch{\"u}hle}, U. and {Shaughnessy}, B. and {Sidher}, S. and {Solanki}, S.~K. and {Speight}, R. and {Spescha}, M. and {Szwec}, N. and {Tamiatto}, C. and {Teriaca}, L. and {Thompson}, W. and {Tosh}, I. and {Tustain}, S. and {Vial}, J. -C. and {Walls}, B. and {Waltham}, N. and {Wimmer-Schweingruber}, R. and {Woodward}, S. and {Young}, P. and {de Groof}, A. and {Pacros}, A. and {Williams}, D. and {M{\"u}ller}, D.},
        title = "{The Solar Orbiter SPICE instrument. An extreme UV imaging spectrometer}",
      journal = {\aap},
         year = 2020,
        month = oct,
       volume = {642},
          eid = {A14},
        pages = {A14},
          doi = {10.1051/0004-6361/201935574},
archivePrefix = {arXiv},
       eprint = {1909.01183},
 primaryClass = {astro-ph.IM},
       adsurl = {https://ui.adsabs.harvard.edu/abs/2020A&A...642A..14S}
}

@article{plowman2023spice,
  title={SPICE point spread function correction: General framework and capability demonstration},
  author={Plowman, JE and Hassler, DM and Auch{\`e}re, F and Cuadrado, R Aznar and Fludra, A and Mandal, S and Peter, H},
  journal={Astronomy \& Astrophysics},
  volume={678},
  pages={A52},
  year={2023},
  publisher={EDP Sciences}
}

@article{dufresne2024chianti,
  title={CHIANTI—an atomic database for emission lines—paper. XVIII. Version 11, advanced ionization equilibrium models: density and charge transfer effects},
  author={Dufresne, RP and Del Zanna, G and Young, PR and Dere, KP and Deliporanidou, E and Barnes, WT and Landi, E},
  journal={The Astrophysical Journal},
  volume={974},
  number={1},
  pages={71},
  year={2024},
  publisher={IOP Publishing}
}

@article{xue2025multiwavelength,
  title={Multiwavelength spectroscopic observations of a quiescent prominence},
  author={Xue, Jianchao and Zhang, Ping and Vial, Jean-Claude and Feng, Li and Zapi{\'o}r, Maciej and Curdt, Werner and Li, Hui and Gan, Weiqun},
  journal={Astronomy \& Astrophysics},
  volume={703},
  pages={A145},
  year={2025},
  publisher={EDP Sciences}
}

@article{zhang2026analysis,
  title={Analysis of Lyman $\beta$ and Lyman $\gamma$ Lines in a Pre-Eruptive and Eruptive Prominence with Solar Orbiter SPICE Observations},
  author={Zhang, Yong and Labrosse, Nicolas and Parenti, Susanna and Kucera, Therese A},
  journal={Monthly Notices of the Royal Astronomical Society},
  pages={stag104},
  year={2026},
  publisher={Oxford University Press}
}




\appendix


\bsp	
\label{lastpage}
\end{document}